\documentclass[%
 reprint,
 amsmath,amssymb,
 aps,natbib
]{revtex4-2}
\usepackage{graphicx}
\usepackage{dcolumn}
\usepackage{bm}
\usepackage{color}
\usepackage{hyperref}
\usepackage{booktabs}
\usepackage{enumitem}

\newcommand{\xzy}[1]{\textcolor{black}{#1}}

\begin{document}


\title{Eccentric Stellar-mass Binary Black Holes: Population, Detectability, and Waveform Analysis in the LISA and LIGO Era}

\author{Zeyuan Xuan}
\email{zeyuan.xuan@physics.ucla.edu}
\affiliation{Department of Physics and Astronomy, UCLA, Los Angeles, CA 90095, USA}
\affiliation{Mani L. Bhaumik Institute for Theoretical Physics, UCLA, Los Angeles, CA 90095, USA}

\author{Smadar Naoz}
\affiliation{Department of Physics and Astronomy, UCLA, Los Angeles, CA 90095, USA}
\affiliation{Mani L. Bhaumik Institute for Theoretical Physics, UCLA, Los Angeles, CA 90095, USA}

\author{Kyle Kremer}
\affiliation{University of California, San Diego, Department of Astronomy \& Astrophysics; La Jolla, CA 92093, USA}

\author{Michael L. Katz}
\affiliation{Jet Propulsion Laboratory, Pasadena, CA 91109, USA}
\affiliation{California Institute of Technology, Pasadena, CA 91125, USA}

\author{Bence Kocsis}
\affiliation{Rudolf Peierls Centre for Theoretical Physics, Parks Road, Oxford OX1 3PU, UK}

\author{Erez Michaely}
\affiliation{Department of Physics and Astronomy, UCLA, Los Angeles, CA 90095, USA}

\date{\today}

\begin{abstract}
Eccentric binary black holes (BBHs) formed through dynamical interactions can significantly contribute to gravitational wave (GW) detections. In this work, we present a simulated catalog of dynamically-formed, stellar-mass BBHs in the local universe, incorporating contributions from the Galactic field (flyby interactions), Galactic nucleus (eccentric Kozai–Lidov evolution), and globular clusters (N-body interactions). \xzy{Our results predict a wide, highly eccentric BBH population in the Milky Way (MW), with source counts of $\sim 25, 9, 3, 1,$ and $<1$ for a 4-year LISA mission ($\sim 36, 13, 5, 2,$ and $1$ for a 10-year extended mission), for $\mathrm{SNR} > 1, 3, 8, 20,$ and $50$, respectively.} Extending this model to cosmological populations, we show that different dynamical channels can produce distinct eccentricity distributions in the LVK band and can contribute hundreds of additional low-SNR mHz sources. Specifically, our model yields a merger rate of $\Gamma \sim 9~\mathrm{Gpc}^{-3}\mathrm{yr}^{-1}$ and \xzy{$\sim 150$} extragalactic mHz BBHs with $\mathrm{SNR} > 1$. However, due to the lower mass and weaker GW signals of stellar-mass BBHs, this number declines sharply at higher detection thresholds (e.g., $\sim 1$ for $\mathrm{SNR} > 8$). We further highlight the impact of eccentric BBH signals on the LISA global fit, showing that their individual harmonics can be independently detected in the Milky Way, and may mimic circular binaries with systematically biased chirp masses. Lastly, we show that post-Newtonian waveforms converge reliably for eccentric BBHs with masses of $\lesssim 10^3\,M_\odot$ in the mHz band. Overall, eccentric BBHs represent a prevalent and promising target for future space-based GW observatories. The simulated catalog and the {\it LISA Eccentricity Astrophysics Package} (\texttt{LEAP}) developed in this work are publicly available at \href{https://github.com/zeyuanxuan/lisa-leap/}{https://github.com/zeyuanxuan/lisa-leap/}. 
\end{abstract}

\maketitle

\section{Introduction}
\label{sec:intro} 

The detection of gravitational waves (GWs) by the LIGO/Virgo/KAGRA (LVK) collaboration has opened a new window beyond traditional electromagnetic observations \citep[see, e.g.,][]{Abbott_2016, 2021arXiv211103634T, Abbott_2023}. Building on this success, space-borne detectors such as LISA, TianQin, and Taiji will probe the mHz GW band in the 2030s, vastly expanding the observable source population \citep[see, e.g.,][]{Amaro_Seoane_2023, luo16, Ruan_2020}. A key distinction between ground-based and space-borne detectors lies in the evolutionary stages they probe: the LVK network observes compact object binaries during their final merger phase, when gravitational radiation has largely circularized their orbits \citep{Peters64}. In contrast, LISA will observe binaries at much earlier stages \citep[see, e.g., ][]{barack04,Mikoczi+12,robson18, chen20gas, amaro+22}, where orbital eccentricity leaves distinct imprints on the waveform and encodes valuable information about their dynamical environments \citep[e.g.,][]{Hoang+19, Fang19, tamanini19, Torres-Orjuela21, Xuan+21, Xuan23acc,Xuan24parameter,xuan2025UCXB}. 

In particular, eccentric stellar-mass binary black holes (BBHs) can have a significant population in the mHz GW band \citep[e.g.,][]{sesana16,Fang_2019}. The existence of such systems is well motivated, as the LVK network has already detected over two hundred stellar-mass BBH mergers, with the latest GWTC-4.0 catalog hinting at their complex dynamical origins \citep{gwtc4}. Notably, current ground-based observations have not yet provided definitive evidence for residual eccentricities in BBH mergers, primarily because gravitational radiation efficiently circularizes orbits by the time binaries enter the high-frequency LIGO band \citep[see, e.g.,][]{Abbott_2019ecc,Lenon_2020, Romero-Shaw_2020,Zevin_2021,gayathri2022eccentricity}. Nevertheless, simulations strongly suggest that a substantial fraction of these observed BBHs might originate from dynamical formation channels. For example, BBHs assembled dynamically in star clusters can undergo frequent external perturbations, including scatterings and GW captures \citep[][]{O'Leary+09,Kocsis_2012,breivik16,rodriguez16,Orazio+18}, which efficiently excite their orbital eccentricities and lead to eccentric mergers \citep{Kremer2019_LISA,Zevin_2019,Samsing+19,Martinez+20,Antonini+19,wintergranic2023binary,Xuan_2025gc}. BBHs in hierarchical triple systems, particularly those near supermassive black holes (SMBH) in galactic nuclei, can undergo significant eccentricity oscillations driven by the eccentric Kozai-Lidov (EKL) mechanism \citep{Kozai1962,Lidov1962,Naoz16}, potentially leading to observable mergers \citep{ Hoang+18,Stephan+19,Randall2019KL,Bub+20,Deme+20,Wang+21,Xuan+23b,knee2024detectinggravitationalwaveburstsblack,grishin2026GWecc}. In addition, fly-by interactions and galactic tides in the field \citep[e.g.,][]{kocsis06,Michaely+19,Michaely+20,Michaely+22,Stegmann_2024}, as well as interactions within active galactic nucleus (AGN) accretion disks \citep{Tagawa+2021,peng2021,Samsing+2022,Munoz+22,Gautham+23}, can also provide significant contributions to the population of eccentric mHz GW sources. These dynamically-formed systems can retain significant, or even extreme, eccentricities at earlier evolutionary stages, resulting in a large underlying mHz GW source population \citep[e.g.,][]{Stephan+19,Hoang+19,Wang+21,Xuan+23b}.

Traditionally, the detectable population of stellar-mass BBHs at cosmological distances was expected to be small, primarily because their signals in the mHz band are relatively weak and often overshadowed by louder, more massive sources such as extreme mass-ratio inspirals (EMRIs) and supermassive black hole binaries \citep[see, e.g.,][]{amaro+22, robson18}. However, recent studies have shown that many dynamically-formed, stellar-mass binaries naturally undergo a wide (semi-major axis $a \gtrsim 0.1\,\mathrm{au}$) and highly eccentric (eccentricity $e \gtrsim 0.9$) progenitor stage before merger \citep[see, e.g.,][]{Xuan+23b,knee2024detectinggravitationalwaveburstsblack}. Although these highly eccentric progenitors are detectable only within short distances ($\sim 10\,\mathrm{Mpc}$), they can remain in the mHz GW band for very long timescales ($\sim 10^5$–$10^7$ years). Such a long observable lifetime leads to a large number of wide, highly eccentric BBHs, which is especially important for the detection of mHz GWs in the very local universe (such as the Milky Way), where proximity naturally compensates for weaker signal strains. As a result, these systems can significantly enlarge the total catalog of stellar-mass GW sources detected in the mHz regime.

However, a self-consistent treatment of astrophysical populations and data analysis strategies for highly eccentric GW sources remains largely unexplored, which creates an urgent need for precise models to unlock LISA’s full scientific potential. In particular, the dominant GW emission from highly eccentric BBHs occurs near pericenter, where the pericenter distance $r_p = a(1-e)$ becomes small \citep{Peters64}. As a result, their emission takes the form of repeated, short-duration bursts \citep[see, e.g., ][]{Kocsis_2012,Tai_2014,Loutrel_2017,Loutrel+20,Romero23,Xuan+23b, Xuan24parameter}. These burst-like features differ qualitatively from the continuous, quasi-circular inspirals targeted by standard GW data analysis methods. Consequently, quasi-circular inspiral templates are both computationally inefficient and morphologically inaccurate for these signals. This mismatch may hinder future detections. For example, the combined unresolved signal from a population of highly eccentric stellar-mass BBHs could form a stochastic gravitational wave background that contaminates the detection of other mHz sources \citep{Xuan24bkg,chen2026implicationslisastochasticsignal}. On the other hand, previous works have conceptually proven that extracting highly eccentric GW signals in mock mHz data analysis is feasible \citep{Xuan24parameter}. Accurately identifying these signals is critical not only for uncovering their dynamical origins, but also for probing their surrounding environments \citep[e.g., ][]{Thompson+11,Hoang+18,Stephan+19,Hoang+20,Naoz+20, Xuan23acc} and improving overall parameter estimation accuracy \citep{Xuan24parameter}.

In this paper, we construct a ready-to-use mock source catalog for eccentric, stellar-mass BBHs, with a particular focus on their orbital parameter distributions and the detectability of their signals in the mHz GW band. The simulated population focuses primarily on the local universe population, where more observational data and simulation results are available, while we also probe the generalization for BBHs at cosmological distances to provide a comprehensive picture. Furthermore, we discuss the validity of post-Newtonian (PN) waveform models for analyzing these eccentric GW sources. Addressing these theoretical and computational challenges is especially important for the development of the LISA global fit pipeline, which requires highly accurate and computationally efficient templates to identify and subtract individual sources from the data stream.

This paper is organized as follows. Section~\ref{sec:burst_properties} outlines the physical properties, orbital timescales, and mHz GW signatures of eccentric BBHs. In Section~\ref{sec:population}, we detail the construction of our mock BBH catalog and explore its astrophysical implications for local and cosmological populations. Section~\ref{sec:globalfit} examines the challenges these multi-harmonic sources pose to the LISA global fit, including potential signal confusion and parameter biases. Section~\ref{sec:waveform} evaluates the convergence and validity limits of post-Newtonian waveform models for analyzing these eccentric systems. Finally, we summarize our results in Section~\ref{sec:discussion}, and provide a guide to the analytical and numerical conventions for eccentric GW signal analysis in Appendix~\ref{sec:appendixa}. The simulated catalog in this work is publicly available via an open-source Python package, the \textit{LISA Eccentricity Astrophysics Package} (\href{https://github.com/zeyuanxuan/lisa-leap/}{\texttt{LEAP}}). This package is also designed for eccentric GW signal analysis, supporting binary orbital evolution, time-domain waveform generation, and SNR estimations in the mHz band.

nless otherwise specified, we set $G=c=1$.

\section{Key Aspects of Eccentric GW Sources in the Millihertz Band}
\label{sec:burst_properties}

\subsection{Basic Properties}\label{subsec:basicproperty}
Due to their proximity, many BBH sources in the local universe can be observed during early evolutionary stages, where they exhibit wide and highly eccentric orbits. Such systems naturally arise as progenitors in a variety of dynamical formation channels. In these scenarios, transient interactions (e.g., fly-bys) or secular processes (e.g., hierarchical triple evolution) can efficiently excite the eccentricity of compact object binaries, leading to strong, ``burst-like'' GW emission during pericenter passages. The intense excitation of GW radiation further decouples the binary system from its surrounding dynamical environment, effectively shrinking and circularizing the orbit until it transitions into a standard, quasi-circular merger. Previous works \citep{Xuan+23b,Xuan24bkg,Xuan_2025gc} have shown that highly eccentric binaries could significantly contribute to the population of stellar-mass GW sources in the mHz band. For completeness, below we briefly summarize the relevant findings.

The lifetime of wide, eccentric compact object binaries is generally longer than that of circular binaries with the same peak GW radiation frequency. This longevity occurs because effective energy emission for a highly eccentric orbit only happens during the pericenter passage, which suppresses the time-averaged energy loss. In particular, for a highly eccentric GW source with component masses $m_1,\,m_2$, its merger timescale can be estimated as \citep[][]{peters63,Xuan+23b}:

\begin{align}
    t_{\mathrm{merger}}\sim & \frac{3}{85m_1m_2M} a^4\left(1-e^2\right)^{7 / 2} 
    \sim  1.12\times10^{7}{\rm yr}\times \nonumber\\&\,\eta_s^{-1}\left(\frac{M}{20\rm M_{\odot}}\right)^{-\frac{5}{3}} \left(\frac{f_{\rm GW}}{1\rm mHz}\right)^{-\frac{8}{3}}\left(\frac{1-e}{0.01}\right)^{-\frac{1}{2}} ,
    \label{eq:lifetime}
\end{align}

where $M=m_1+m_2$, $\eta_s=4 m_1m_2/(m_1+m_2)^{2}=4q/(1+q)^2$ is unity for equal-mass sources, with $q=m_1/m_2$ denoting the mass ratio, and $f_{\rm GW}$ denotes the characteristic peak GW frequency associated with the binary's pericenter passage:
\begin{equation}
f_{\rm GW} \sim 2f_{\rm orb}(1-e)^{-3/2},
\label{eq:peakf}
\end{equation}
where $f_{\rm orb}$ is the orbital frequency. We adopt this convention because it smoothly reduces to the circular-orbit GW frequency in the limit $e\rightarrow0$. An alternative definition commonly adopted in the literature is $f_{\rm peak}=f_{\rm orb}(1+e)^{1/2}(1-e)^{-3/2}$ \citep{O'Leary+09} \citep[see also][for more details]{Hamers2021fpeak}; for highly eccentric systems, the two definitions differ by a factor of $\sqrt{2}$.

Note that the merger timescale in Equation~(\ref{eq:lifetime}) is much longer than that of a circular inspiral with the same $f_{\rm GW}$ \citep{Xuan+23b}:

\begin{align}
t_{\rm inspiral} \sim 5.5\times10^{4} {\rm yr} \,\eta_s^{-1} \left(\frac{M}{20 \,\rm M_{\odot}}\right)^{-\frac{5}{3}}\left(\frac{f_{\rm GW}}{1\,\rm mHz}\right)^{-\frac{8}{3}}
\ ,
\label{eq:inspiral}
\end{align}

The longer lifetime of eccentric progenitor systems indicates a larger population expectation in the universe, which yields an advantage for detecting these systems. 

\begin{figure}
    \centering
    \includegraphics[width=3.3in]{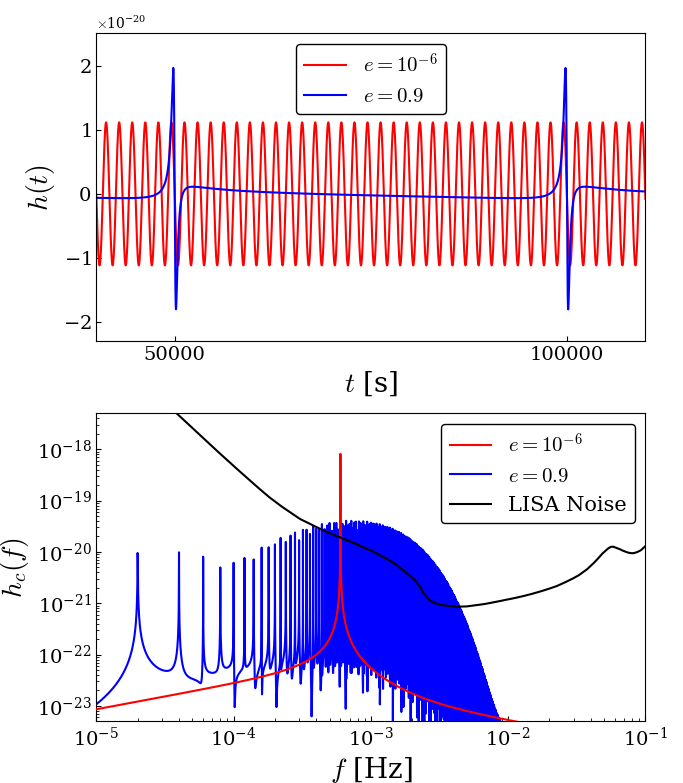}
    \caption{{\bf{GW waveforms and corresponding characteristic strains for BBHs with different eccentricities.}}
    We show the GW signals (plus polarization) from BBH systems with 
    $m_{1}=25$~M$_{\odot},$
    $m_{2}=20$~M$_{\odot}$, placed at $8$~kpc and observed for $0.5$~yrs. Blue curves represent a highly eccentric BBH with $f_{\rm orb}=2\times10^{-5}$~Hz, $e=0.9$, Red curves represent a quasi-circular BBH with $f_{\rm orb}=3\times10^{-4}$~Hz, $e=10^{-6}$. These parameters are chosen such that both BBHs will have
     $\rm SNR\sim 120$ for a 0.5-year observation. {\it Upper Panel} shows a zoom-in view of their time-domain waveforms, $h(t)$; {\it Bottom Panel} shows their characteristic strains ({\it numerical spectrum}, $h_{\rm c,num}$, see Appendix~\ref{app:plothighe} for details), compared with the LISA noise curve ($\sqrt{fS_n(f)}$, black line). The signals are numerically generated using a post-Newtonian waveform model (see Section~\ref{sec:waveform}).}
    \label{fig:burst_waveform}
\end{figure}

On the other hand, highly eccentric GW signals have unique time and frequency patterns. For example, Figure~\ref{fig:burst_waveform} shows a comparison between GWs from circular and highly eccentric BBHs. As shown, the GW signal from eccentric systems (blue lines) is radiated primarily during their periapsis passages, rather than continuously throughout the orbit. Therefore, eccentric signals in GW data analysis can appear as a sequence of {\it repeated bursts}, with the time interval between each burst equal $1/f_{\rm orb}$ (see the {\it Upper Panel}) and radiated power distributed over a wide range of harmonic frequencies (see the {\it Bottom Panel}). 

The unique waveform patterns of eccentric binaries present challenges for realistic data analysis. Nevertheless, a robust estimate of their detectability can be obtained by computing the total signal-to-noise ratio through summing the energy contributions from all individual frequency harmonics \citep[e.g.,][]{O'Leary+09,Kocsis_2012,Hoang+19,Xuan+23b}. Furthermore, for sources with slow orbital evolution (i.e., $t_{\rm merger}$ much longer than the observation time $T_{\rm obs}$), their SNR can be analytically approximated as:
\begin{equation}
{\rm SNR} 
\sim \frac{h_{\rm burst}}{\sqrt{S_n\left(f_{\rm GW }\right)}} \sqrt{T_{\rm o b s}\left(1-e\right)^{3 / 2}}\,, 
\label{eq:snrnew}
\end{equation}
in which $S_{\mathrm{n}}(f)$ is the
spectral noise density of LISA evaluated at GW frequency $f$
\citep[][]{Klein+16,Robson+19}. The subscript ``n'' in $S_n(f)$ stands for ``noise'', which is different from the harmonic index $n$ used throughout this paper. $h_{\rm burst}$ is the strain amplitude of a GW burst during the binary's pericenter passage \citep{peters63,Kocsis_2012}: 
\begin{align}
    &h_{\rm burst} \sim \sqrt{\frac{32}{5}}\frac{m_1m_2}{D_la(1-e)}\nonumber\\
    &\sim  3.49\times 10^{-21}\eta_s \left(\frac{M}{20\,\rm M_{\odot}}\right)^{\frac{5}{3}} \left(\frac{f_{\rm GW}}{1\,\rm mHz}\right)^{\frac{2}{3}}\left(\frac{D_{l}}{8\,\rm kpc}\right)^{-1}\ ,
    \label{eq:amplitude}
\end{align}
where $D_{l}$ is the luminosity distance of the source, and the constant coefficient $\sqrt{32/5}$ is due to the average of the GW strain amplitude over the binary's orientation and sky location. 

Additionally, if a wide eccentric binary has an orbital period longer than the observation time, its GW emission may not occur within the entire observation window. In this case, the SNR estimate in Equation~(\ref{eq:snrnew}) can be written as the product of the probability of detecting a burst during the observation window, $P_{\rm burst}=f_{\rm orb}T_{\rm obs}$, and the SNR of a single GW burst:

\begin{equation}
    {\rm SNR_{singleburst}} \sim \dfrac{h_{\rm burst}}{\sqrt{f_{\rm GW} S_n(f_{\rm GW })}}\, .
    \label{eq:singleburst}
\end{equation}

We note that Equations~(\ref{eq:snrnew})--(\ref{eq:singleburst}) provide quick analytical estimates of the SNR. However, for a more rigorous treatment, we adopt the full summation over GW harmonic power to compute the SNR throughout this work (see Equation~(\ref{eq:snrsum3}) in Appendix~\ref{sec:appendixa}). These two approaches generally agree within a factor of $\sim 2$, as shown in Figure~\ref{fig:pn_accuracy3}.

\xzy{In addition, the detection and parameter estimation of eccentric binaries differ substantially from those of quasi-circular systems. In particular, source confusion and parameter degeneracies in the LISA global fit may complicate their identification, especially at low signal-to-noise ratios. Nevertheless, previous works have shown that these systems can be effectively recovered with appropriate eccentric waveform templates in mock data analyses \citep{Xuan24parameter}. Furthermore, their multiple harmonic signatures and orbital precession patterns can provide additional observables that help break degeneracies and improve source identification and parameter estimation \citep{Xuan_2025gc,Xuan23acc}. Therefore, throughout this paper, detectable source populations are reported using a threshold of $\mathrm{SNR}>8$ for the nominal 4-year LISA mission \citep{Amaro_Seoane_2023}, while corresponding results for a 10-year extended mission are also included for comparison.}


In Table~\ref{tab:typical_params}, we summarize the typical orbital properties, lifetimes, and detectable distances of stellar-mass black hole binaries at different evolutionary stages. As shown in the table, highly eccentric progenitors of BBH mergers are typically detectable out to $\sim10$ -- $10^4$~kpc, enabling LISA to detect such systems across the local universe. This detection range, combined with their long lifetimes and potentially large population, makes highly eccentric BBHs promising targets for mHz GW observations in the local universe, and motivates studying their population in the local universe (see Section~\ref{subsec:catalog}).

\subsection{Characteristic Strain Representations}\label{subsec:hcreps}

As discussed above, the waveform and spectral structure of highly eccentric GW sources differ qualitatively from those of quasi-circular binaries. As a result, there are different ways in the literature to plot their characteristic strain spectrum \citep[e.g.,][]{O'Leary+09,Kocsis_2012,Orazio+18,Hoang+19,Kremer2019_LISA,Emami+20,Deme+20,Xuan+23b,Naoz+23}. For completeness, we summarize these approaches here; see Appendix~\ref{sec:appendixa} for further details.

We begin by recalling the standard definition of the characteristic strain for a quasi-circular source \citep[see, e.g., ][]{Finn_2000,Moore_2014, Kupfer_2018,Tauris_2018,Robson+19}:
\begin{equation}\label{eq:hcmixed0}
    h_c = \sqrt{2} h_0 \sqrt{ \min\left\{ \frac{f^2}{\dot{f}}, \, f T_{\rm obs} \right\} } \ ,
\end{equation}
in which $h_0 = \sqrt{32/5} \,m_1 m_2/(D_l a)$ is the root-mean-square (rms) time-domain GW amplitude, $f \sim 2 f_{\rm orb}$ is the GW frequency for a quasi-circular binary (corresponding to the dominant second harmonic), and $\dot{f}$ is the frequency shift rate driven by GW emission (chirp rate). 

The two regimes in the minimum function of Equation~(\ref{eq:hcmixed0}) differ fundamentally in their physical meaning, in the $f^2/\dot{f}$ regime, the frequency of the source evolves rapidly during $T_{\rm obs}$, and the {\it integrated area} between $h_c=\sqrt{2} h_0 \sqrt{ f^2/\dot{f}}$ contour and detector noise curve in the characteristic strain plot represents the SNR of the corresponding GW harmonic \citep{Finn_2000, Moore_2014}; in the $f T_{\rm obs}$ regime, the source evolves slowly, and the {\it height} of $h_c=\sqrt{2} h_0 \sqrt{  f T_{\rm obs}}$ in the characteristic strain plot directly reflects the SNR of signal \citep{Kupfer_2018,Tauris_2018,Robson+19}.

\begin{figure}
    \centering
\includegraphics[width=3.4in]{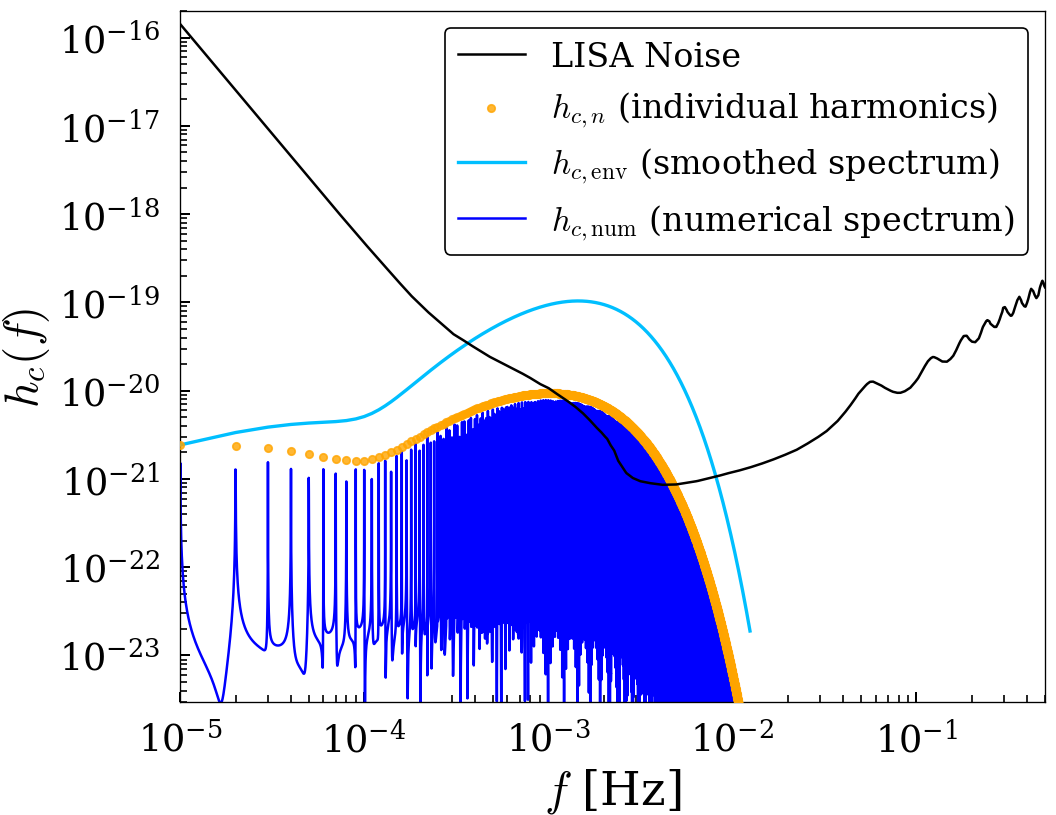}
\caption{\textbf{Comparison of three different characteristic strain representations for the same eccentric BBH source.} The plot displays the individual harmonic representation ($h_{c,n}$, scatter points), the smoothed spectrum representation ($h_{c,\rm env}$, solid envelope), and the numerical spectrum representation ($h_{c,\rm num}$, oscillatory curve) for the characteristic strain of a BBH system with $m_1 = 15 M_\odot$, $m_2 = 10 M_\odot$, $f_{\rm orb} = 10^{-5}$~Hz, $e = 0.95$, $T_{\rm obs}=0.5$~yr, and $D_l = 8$~kpc. In the non-evolving limit ($\dot{f} \to 0$), the peak heights of the numerical spectrum coincide with the individual harmonics, and differ from the smoothed spectrum envelope by a factor of $\sqrt{f_{\rm orb}/f}$. This system has an overall SNR of $\sim 60$. }\label{fig:hc_representations}
\end{figure}

For an eccentric binary, the GW signal is made up of multiple harmonics (see, e.g., Figure~\ref{fig:burst_waveform}), therefore, a natural way to plot its $h_c$ is to treat each harmonic component as an individual ``quasi-circular" source, and plot their $h_c$ following Equation~(\ref{eq:hcmixed0}) \citep[see, e.g., ][]{O'Leary+09,Kocsis_2012,Orazio+18,Kremer2019_LISA,Emami+20}:
\begin{equation}\label{eq:hcn,0}
    h_{c,n} = \sqrt{2} h_n \sqrt{ \min\left\{ \frac{f_n^2}{\dot{f}_n}, \, f_n T_{\rm obs} \right\} } \ ,
\end{equation}
in which $h_n$ is the rms time domain amplitude of the n-th harmonic (Equation~\ref{eq:hntime}), $f_n=nf_{\rm orb}$, and $\dot{f}_n=n\dot{f}_{\rm orb}$ can be directly related to the eccentric binary's orbital evolution. This approach, which we refer to as the {\it individual harmonic representation}, plots each harmonic independently. The height of each harmonic reflects its {\it individual} detectability  (see, e.g., orange points in Figure~\ref{fig:hc_representations}).


However, for highly eccentric sources, the number of harmonics is large, and their number density varies depending on binary orbital frequency. As a result, plotting individual harmonics can become overwhelmingly cluttered, and their height does not directly reflect the {\it overall} detectability of an eccentric source. This motivates the second approach in the literature to plot the $h_c$ envelope for an eccentric GW source, which we refer to as the {\it smoothed spectrum representation} \citep[see, e.g., ][]{Hoang+19,Deme+20,Xuan+23b,Naoz+23}:
\begin{equation}\label{eq:hc_highe0}
h_{c,\rm env}(f) = \sqrt{2}\, h_n(f)\, \sqrt{\frac{f^2 T_{\rm obs}}{f_{\rm orb}}} \ ,
\end{equation}
where $h_n(f)$ represents the interpolated rms time-domain strain amplitude as a continuous function of frequency, constructed from the discrete harmonic values $h_n$, and satisfies $h_n(f_n) = h_n$ at each harmonic frequency $f_n$ (see Equation~(\ref{eq:hc_highe}) for details). 

Note that Equation~(\ref{eq:hc_highe0}) assumes the source has negligible orbital evolution during $T_{\rm obs}$. For the general evolving cases, the contour of Equation~(\ref{eq:hc_highe0}) can shift in the characteristic strain plot \citep[see, e.g., fig.1 in][]{Hoang+19}, and the overall spectrum can be computed via an integral of $h^2_{c,\rm env}(f)$ over $dT_{\rm obs}$. For more details, see Appendix~\ref{app:plothighe}.

The definition in Equation~(\ref{eq:hc_highe0}) is constructed to capture the correct spectral contribution of a single eccentric binary across frequencies, such that the standard characteristic strain integral yields the correct accumulated SNR (see Equation~\ref{eq:snrintegral}). In this representation, the area between the envelope and the detector noise curve provides a visually intuitive estimate of the total SNR contribution (see, e.g., the cyan line in Figure~\ref{fig:hc_representations}). 

We note that $h_{c,\rm env}$ does not correspond to the geometric upper envelope of individual harmonics $h_{c,n}$. For example, Figure~\ref{fig:hc_representations} compares the {\it individual harmonic representation}, $h_{c,n}$; the {\it smoothed spectrum representation}, $h_{c,\rm env}$; and the {\it numerical spectrum}, $h_{c,\rm num}$, for a slowly evolving BBH system with $m_1 = 15\,M_\odot$, $m_2 = 10\,M_\odot$, $e = 0.95$, $f_{\rm orb} = 10^{-5}\,\mathrm{Hz}$, $D_l = 8\,\mathrm{kpc}$, and $T_{\rm obs} = 0.5\,\mathrm{yr}$. The numerical spectrum $h_{c,\rm num}$ is obtained by numerically generating the binary's time-domain waveform, computing its Fast Fourier Transform, and normalizing the result (see Appendix~\ref{app:hcnumerical} for details).

As shown in the figure, the discrete amplitudes of individual harmonics ($h_{c,n}$; Equation~\ref{eq:hcn,0}) closely match the peak heights in the numerical spectrum ($h_{c,\rm num}$; Equation~\ref{eq:hnc_num}). Small differences arise because $h_{c,n}$ represents the sky-averaged GW amplitude, whereas $h_{c,\rm num}$ in this example is computed using only the plus polarization and with fixed observation angles. In addition, when the harmonic frequencies $n f_{\rm orb}$ do not coincide exactly with the discrete FFT frequency grid (spaced by $1/T_{\rm obs}$), the finite-window effect causes the sampled peak in $h_{c,\rm num}$ to fall slightly below the true harmonic amplitude, with the ``leaked'' power redistributed into adjacent bins; this redistribution preserves the total SNR by Parseval's theorem, but can lower individual peak heights by up to a factor of $2/\pi$ in the worst case of bin misalignment.

Furthermore, both $h_{c,n}$ and $h_{c,\rm num}$ lie below the smoothed spectral envelope ($h_{c,\rm env}$; Equation~\ref{eq:hc_highe0}). This offset reflects the fact that $h_{c,\rm env}$ renormalizes the power of discrete harmonics to preserve the integrated SNR, whereas the harmonic number density in the $h_{c,n}$ representation is not uniform in logarithmic frequency space and depends on the binary parameters. As a result, for eccentric sources satisfying $\dot{f}_{\rm orb}<f_{\rm orb}/T_{\rm obs}$, the individual harmonic amplitude at the $n$-th harmonic is lower than the smoothed envelope by a factor of $\sqrt{f_{\rm orb}/f}$ (see Equations~\ref{eq:hcn,0} and \ref{eq:hc_highe0}):
\begin{equation}\label{eq:hcdifference}
    h_{c,n} = \sqrt{2} h_n \sqrt{ f_n T_{\rm obs}} =  h_{c,\rm env} (f_n) \sqrt{\frac{f_{\rm orb}}{f_n}}\ .
\end{equation}

\begin{table*}[htbp]
\centering
\small
\setlength{\tabcolsep}{4pt}
\caption{Typical Orbital Parameters and Lifetimes of Dynamically-Formed, Stellar-mass BBHs at Different Evolutionary Stages \label{tab:typical_params}}
\begin{tabular}{cccccc}
\toprule
Evolutionary Stage & $a$ & $e$ & $t_{\rm merger}$ & $D_{l,\rm max}$ & GW Band \\
\midrule
Highly Eccentric Progenitor & $\sim 10^{-1} - 10^2\,\mathrm{au}$ & $\gtrsim 0.9 $ & $\sim 10^5 - 10^8\,\mathrm{yr}$ & $10\,\mathrm{kpc} - 10\,\mathrm{Mpc}$ & mHz \\
Quasi-circular Inspiral     & $\sim 10^{-3}\,\mathrm{au}$        & Moderate       & $\sim 10^2 - 10^5\,\mathrm{yr}$ & $\sim 500\,\mathrm{Mpc}$ & mHz \\
Merger                      & $\sim 10^{-6}\,\mathrm{au}$        & $\sim 0$       & $\lesssim 1\,\rm s $          & $\sim 1 - 10\,\mathrm{Gpc}$ & $\sim 100\,\mathrm{Hz}$ \\
\bottomrule
\end{tabular}

\vspace{1ex}
{\raggedright \footnotesize \textit{Note:} Summary of typical orbital parameters, remaining lifetimes ($t_{\rm merger}$), maximum detectable luminosity distances ($D_{l,\rm max}$), and primary gravitational wave observational bands for dynamically formed stellar-mass BBHs. The table illustrates the evolution from wide, highly eccentric progenitor binaries down to the final merger phase \citep{sesana16, abbott20GW190521}. Note that exact merger timescales and maximum detectable distances are highly dependent on the total system mass; the bounds presented here are estimated using a fiducial mass range of $10 - 100\, M_{\odot}$. The mHz GW band detectability is estimated for LISA, whereas the $\sim 100\,$Hz band represents the current ground-based detector networks like LIGO-Virgo-KAGRA.\par}
\end{table*}

\section{Simulated Eccentric BBH Population }
\label{sec:population}
\subsection{Creating a mock BBH Catalog}
\label{subsec:catalog}

We first assemble a realistic sample of dynamically-formed BBHs in the Milky Way. In particular, we adopt the models developed by \citet{Xuan+23b} (see their Appendix A) and \citet{Xuan_2025gc} (see their Section 2). This suite of models combines multiple dynamical formation channels that are theoretically expected to produce a significant number of GW sources in different regions of the galaxy, and involves the recent simulation results and observational constraints on the corresponding compact object populations. Specifically:

\begin{description}[leftmargin=0pt, style=sameline]
    \item[Galactic Field] We model the population of fly-by-induced highly eccentric GW sources in the Galactic field, following the approach in \citet{Michaely+19,Michaely+22,Michaely+20}. These sources originate from wide binary black holes ($ a\sim 10^2  - 10^4$\,au) residing in the Milky Way disk, with their initial spatial distribution assumed to follow the Galactic stellar density profile (see eqs.~23 – 26 of \citet{Michaely+19}). For simplicity, we set a fixed fly-by perturber mass of $0.6\,M_\odot$ and adopt a stellar velocity dispersion of $50\,\mathrm{km\,s^{-1}}$ for the Galactic disk. The initial BBH mass follows the standard LIGO-inferred distribution \citep{PhysRevX.13.011048}, and is set to form in a starburst 10\,Gyr ago, with a wide BBH fraction $f_{\rm BBH}=7\times10^{-4}$ relative to the total stellar population \citep{Michaely+19}. We then evolve this initial sample to the present day, including the effect of depletion due to binary evaporation and fly-by induced mergers. The resulting present-day merger rate is $\sim 3\times 10^{-7}\,\rm yr^{-1}$ in the Milky Way, and translates to a volumetric rate of $\sim 3~\mathrm{Gpc}^{-3}\,\mathrm{yr}^{-1}$ in the Universe, assuming a galaxy number density of $0.01~\rm Mpc^{-3}$ \citep{Conselice_2005, van_Dokkum_2013}.
 
    \item[Galactic Nucleus (GN)] We model BBH mergers in Galactic nucleus following \citet{Hoang+18,Hoang+19,Xuan+23b}. Specifically, the Galactic center SMBH can secularly perturb its surrounding stellar-mass binaries through hierarchical triple evolution \citep[eccentric Kozai–Lidov mechanism; see, e.g.,][]{Kozai1962,Lidov1962, Naoz16}, resulting in eccentricity excitation and subsequent GW mergers. We model this hierarchical triple evolution using secular equations up to the octupole level of approximation \citep{Naoz+13}, including general relativistic precession \citep{naoz13} and GW emission \citep{Peters64}. The initial orientations of stellar-mass black hole binaries are set to be isotropic, and systems that do not satisfy the hierarchical stability criteria at formation are excluded \citep{Naoz+13,Naoz16}.
    
    Within this framework, the BBH population is further divided into two sub-components (see Appendix A of \citet{Xuan+23b} for details):
    \begin{itemize}
    
    \item \textit{Main Population:}
    The main stellar population of the Galactic center nuclear star cluster is expected to lie within $\sim 5~\rm pc$ from the SMBH, with ages of $\sim 2$--$8~\rm Gyr$ and a total stellar mass of $\sim 1.8\times 10^{7}$~M$_{\odot}$ \citep[e.g.,][]{Pfuhl11,Launhardt02}. We assume BBHs form in this population with a continuous steady-state distribution, with an effective replenishment rate of $\Gamma_{\rm rep}\sim 3\times 10^{-6}~\rm yr^{-1}$, and with masses following the LIGO-inferred distribution. The spatial distribution of systems follows a radial density profile scaling as $\rho \propto r^{-2}$ toward the Galactic center. We note that under the steady-state approximation, the total replenishment rate $\Gamma_{\rm rep}$ equals the total depletion rate due to BBH mergers or evaporation, and is therefore larger than the actual GW merger rate. This configuration results in a simulated BBH merger rate of $2\times 10^{-7}\,\rm yr^{-1}$ in the Milky Way, or $\sim 2~\mathrm{Gpc}^{-3}\,\mathrm{yr}^{-1}$ in the Universe, assuming a galaxy number density of $0.01~\rm Mpc^{-3}$. 
    
    \item \textit{Young Nuclear Cluster (YNC) Population:}
    Observations have shown the presence of a young nuclear star cluster in the Milky Way center \citep{Paumard06,Lu2013}, located within $\sim 0.5~\rm pc$ from the SMBH, with ages of $\sim 2$ -- $8~\rm Myr$ and a total mass of $\sim 1.4$ -- $3.7\times 10^{4}~M_{\odot}$. To simulate the corresponding BBH population, we assume that this YNC was formed in a starburst $\sim 2$ -- $8~\rm Myr$ ago, with a top-heavy mass distribution and all massive stars residing in binary systems \citep{2014ApJ...782..101P}. The massive main-sequence binaries are then evolved into their compact object stage, with remnant black hole masses following \citet{woosley02,Belczynski_2020}. According to the simulation results, $\sim 100$--$400$ BBHs are formed in the YNC (assuming no natal kicks), with most of these black holes having masses of $\sim 10~M_{\odot}$ \citep[see, e.g.,][]{Bond84,Fryer01}. For conservative purposes, we adopt a number of 100 newly formed BBHs and evolve them using the same hierarchical triple simulations. Since this model is time-evolving, it does not correspond to a fixed LIGO merger rate; however, it is expected to contribute $\sim 1$ -- $4$ BBHs detectable by LISA  (see fig.~7 of \citet{Xuan+23b}).
    
    \end{itemize}
\begin{table*}[htbp]
\centering
\setlength{\tabcolsep}{10pt}
\caption{Expected Number of Dynamically-formed BBHs in the MW, for Different SNR Thresholds. \label{tab:snr_stats}}
\begin{tabular}{cccccc}
\toprule
SNR & $T_\mathrm{obs}$ & Total & Field & GN & GCs \\
\midrule
$>50$ & 4\,yr  & 0.39 (0.65) & 0.10 (0.30) & 0.15 (0.36) & 0.14 (0.42) \\
      & 10\,yr & 1.04 (0.97) & 0.17 (0.43) & 0.44 (0.64) & 0.43 (0.62) \\
\midrule
$>20$ & 4\,yr  & 1.45 (1.31) & 0.31 (0.58) & 0.57 (0.76) & 0.57 (0.83) \\
      & 10\,yr & 2.26 (1.34) & 0.36 (0.57) & 1.02 (0.88) & 0.88 (0.87) \\
\midrule
$>8$  & 4\,yr  & 3.04 (1.77) & 0.61 (0.76) & 1.37 (1.23) & 1.06 (1.12) \\
      & 10\,yr & 4.74 (1.98) & 0.93 (0.80) & 2.20 (1.27) & 1.61 (1.30) \\
\midrule
$>3$  & 4\,yr  &  8.52 (2.96) & 1.39 (1.14) & 3.46 (1.84) &  3.67 (1.86) \\
      & 10\,yr & 12.95 (3.47) & 2.23 (1.22) & 5.46 (2.38) &  5.26 (2.30) \\
\midrule
$>1$  & 4\,yr  & 24.65 (5.17) & 3.45 (1.76) & 10.00 (3.50) & 11.20 (3.32) \\
      & 10\,yr & 36.38 (5.13) & 4.51 (1.98) & 15.99 (3.37) & 15.88 (4.06) \\
\bottomrule
\end{tabular}

\vspace{1ex}
{\raggedright \small \textit{Note:} \xzy{Number of simulated eccentric BBHs within the Milky Way that are above different SNR thresholds, for the nominal 4-year and the extended 10-year LISA mission durations. The primary value in each cell represents the expected number of observable systems, while the value in parentheses indicates the $1\sigma$ standard deviation derived from the simulated population. The total expected counts are subdivided by their local environment: the Galactic field (Field), the Galactic nucleus (GN), and globular clusters (GCs).}\par}
\end{table*}

    \item[Globular Clusters (GCs)] The GC population is drawn from the \texttt{CMC} (Cluster Monte Carlo) Catalog of \citet{Kremer_2020}, which is based on Monte Carlo $N$-body simulations, and later calibrated to reproduce the observed properties of Milky Way globular clusters \citep{Xuan_2025gc}. These simulations self-consistently follow the dynamical evolution and stellar evolution of dense star clusters, and track the formation and evolution of BBHs within them, including their masses, semi-major axes, and eccentricities across cosmic time.

    To construct a Galactic GC population, each observed Milky Way GC is associated with a best-fit \texttt{CMC} model based on its present-day global properties (e.g., cluster mass, metallicity, and Galactocentric distance). From the corresponding model, we sample cluster snapshots at late evolutionary times ($\sim 8$--$13.7~\rm Gyr$) to account for uncertainties in cluster ages, and extract the full BBH population present at that epoch. This procedure is repeated to generate multiple realizations of the Galactic GC system. For each BBH identified in these clusters, we compute its GW properties using the orbital parameters directly provided by the \texttt{CMC} simulations. The source distance adopted for SNR calculations is taken to be the present-day heliocentric distance of its host GC. For BBHs that have been dynamically ejected from their parent clusters, we also approximate their distances using the current location of the original host GC. Our model results in a simulated BBH merger rate of $4\times 10^{-7}\,\rm yr^{-1}$ in the Milky Way, or $\sim 4~\mathrm{Gpc}^{-3}\,\mathrm{yr}^{-1}$ in the Universe, assuming a galaxy number density of $0.01~\rm Mpc^{-3}$.
\end{description}



Table~\ref{tab:snr_stats} shows the expected number of detectable systems originating from the aforementioned dynamical channels. The projected SNR of the systems is estimated based on their simulated orbital parameters and spatial distances within the mock catalog using Equation~(\ref{eq:snrsum5}), assuming a 4-year (10-year) LISA observation period, respectively. The numbers enclosed in brackets indicate the statistical uncertainty (standard deviation) inherent to the model, arising primarily from the finite sampling size of the numerical simulations. As shown by the table, dynamical channels alone can contribute a highly significant number of detectable systems, even when restricting our analysis to the Milky Way's population. Specifically, \xzy{for a 4-year (10-year) LISA observation, approximately $25, 8.5, 3.0, 1.5, 0.4$ ($36, 13, 4.7, 2.3, 1.0$) dynamically-formed MW BBHs can exceed SNR thresholds of $1, 3, 8, 20, 50$, respectively.} This result is consistent with our previous theoretical frameworks \citep[][]{Xuan+23b,Xuan_2025gc}, and demonstrates that these bursting systems represent promising targets for future data analysis pipelines.

\begin{figure*}
    \centering
    \includegraphics[width=5.5in]{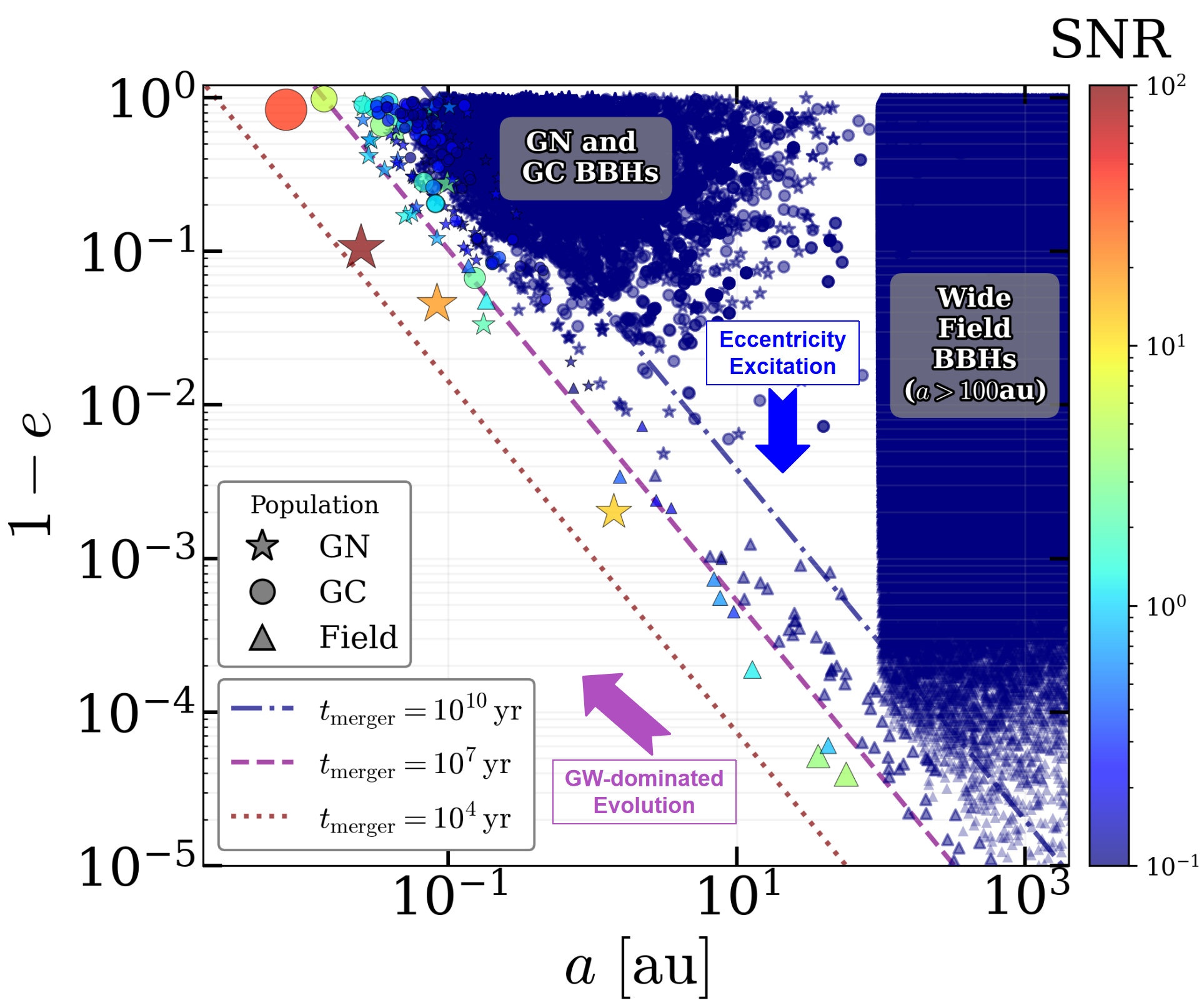}
\caption{\textbf{Dynamically-formed BBHs in a mock realization of the Milky Way.} 
We plot the orbital parameter distribution $(a, 1-e)$ of simulated BBHs in a mock realization of the Milky Way. Different formation channels are indicated by marker shapes: Galactic nucleus (GN; stars), globular clusters (GC; circles), and the Galactic field (triangles). \xzy{Marker colors and sizes represent the expected GW signal-to-noise ratio, assuming a 4-year LISA observation.} Diagonal dashed and dotted lines show contours of constant merger timescale ($t_{\rm merger} = 10^{10}, 10^7$, and $10^4\,\mathrm{yr}$) for an example BBH mass of $10\,M_\odot + 10\,M_\odot$. We further illustrate the evolutionary pathways using arrows. In particular, dynamical perturbations excite eccentricity (blue arrow) until GW emission becomes strong enough to decouple the system from the environment ($t_{\rm merger} \sim 10^{4}-10^{10}\,\mathrm{yr}$), which further shrinks and circularizes the orbit (purple arrow). We note that a cutoff at $a = 100\,\mathrm{au}$ is artificially imposed on the wide field population for conservative purposes. In addition, most of the detectable BBH systems cluster near the $t_{\rm merger} \sim 10^7\,\mathrm{yr}$ contour, which is because contours of constant signal-to-noise ratio (e.g., $\rm SNR=8$) are approximately parallel to those of constant $t_{\rm merger}$ in this regime, see Sections~\ref{subsec:catalog} and \ref{subsec:astro} for details.}
    \label{fig:mock_mw}
\end{figure*}

\xzy{We note that this catalog is not complete, and the BBH source numbers presented here should be regarded as conservative estimates.} In particular, this catalog includes the full BBH populations from the three major dynamical formation channels: fly-by mergers in the Galactic field, EKL-driven mergers in the Galactic Center, and dynamically assembled mergers from GC N-body simulations; \xzy{however, the overall BBH population in the Milky Way likely includes contributions from other pathways not modeled here, notably the isolated binary evolution channel \citep[e.g.,][]{Mandel_2016, breivik16} and alternative dynamical mechanisms such as hierarchical field triples. Including these additional populations would likely increase the number of detectable stellar-mass BBHs in LISA.} Furthermore, the assumptions of each model have inherent uncertainties; for example, the exact density profiles and spatial orientation distributions of BBHs in the Galactic nucleus remain highly uncertain, and the fraction of wide BBHs in the Galactic Field is poorly constrained due to the lack of observation. However, this model captures the major dynamical formation channels as proposed by current theoretical studies, and involves the most up-to-date simulation results and observational constraints on compact object populations. Furthermore, each individual channel yields merger rates that are consistent with the LIGO-inferred rate \citep[$14$–$26$~Gpc$^{-3}$~yr$^{-1}$;][]{LIGO2025}. In future LISA observations, targeted searches for these specific sources can, in turn, constrain the underlying model assumptions and population properties.

Figure~\ref{fig:mock_mw} shows a mock realization of the Milky Way BBH population from our simulation. We plot the semimajor axis $a$ and eccentricity (as $1-e$) for each BBH system in the mock catalog. Galactic nucleus BBHs are shown as stars, globular cluster BBHs as circles, and Galactic field BBHs as triangles. The expected GW SNR is mapped to the color of each point, \xzy{assuming a 4-year LISA observation.} For visual clarity, we also enlarge the size of the systems with higher SNR, and overplot contours of equal merger timescale, corresponding to $t_{\rm merger}=10^4$, $10^7$, and $10^{10}$ years for a representative BBH mass of $10\,M_\odot+10\,M_\odot$. We note that the wide field BBH population exhibits a sharp boundary at $a=100$\,au, which is caused by our conservative modeling assumptions. The existence of BBHs with tighter initial separations remains highly uncertain due to the common-envelope evolution; therefore, systems with $a<100$\,au are excluded.

As shown by the figure, detectable BBH systems span a variety of orbital parameters, with a substantial fraction having significant or even extreme eccentricities upon entering the mHz band. This phenomenon is a direct physical consequence of their dynamical formation nature. As discussed in the Introduction, dynamically-formed BBHs are more likely to be born from a wide orbital configuration (see the top-middle and top-right corners of Figure~\ref{fig:mock_mw}). Environmental perturbations excite their eccentricity (driving the population downwards in Figure~\ref{fig:mock_mw}), until the gravitational wave radiation becomes sufficiently dominant to decouple the binary system from further eccentricity increase (see, e.g., the parameter region where $t_{\rm merger}\lesssim 10^{7}$ yrs). Once decoupled, the systems can further shrink and circularize until becoming a standard merger event (migrating towards the top-left corner, as indicated by the purple arrow). 

In addition, Figure~\ref{fig:mock_mw} shows that the observable mHz BBH population in the Milky Way is dominated by long-living systems, with a significant fraction being wide and highly eccentric. This is a natural consequence of the fact that the expected number of systems in a specific evolutionary stage is proportional to the lifetime of that stage. Interestingly, BBHs located in the Milky Way typically cross the detection threshold of LISA when their remaining merger timescale reaches $\sim 10^7$ years (thus, we expect the majority of the population to cluster here). This coincidence arises because the analytical contours defining constant $t_{\rm merger}$ (${\rm log}_{10}a+0.88{\rm log}_{10}(1-e)\sim \rm constant$) is nearly parallel to the lines defining constant observational SNR (${\rm log}_{10}a+0.83{\rm log}_{10}(1-e)\sim \rm constant$). This parallel alignment indicates that reaching a specific SNR level roughly corresponds to reaching a specific remaining merger timescale across the parameter space (for a detailed derivation, see \citet[][]{Xuan+23b}).

The mock catalog used in this work is publicly available at \url{https://github.com/zeyuanxuan/lisa-leap/}.

\subsection{Astrophysical Implication}
\label{subsec:astro}
\begin{figure*}
    \centering
    \includegraphics[width=7.2in]{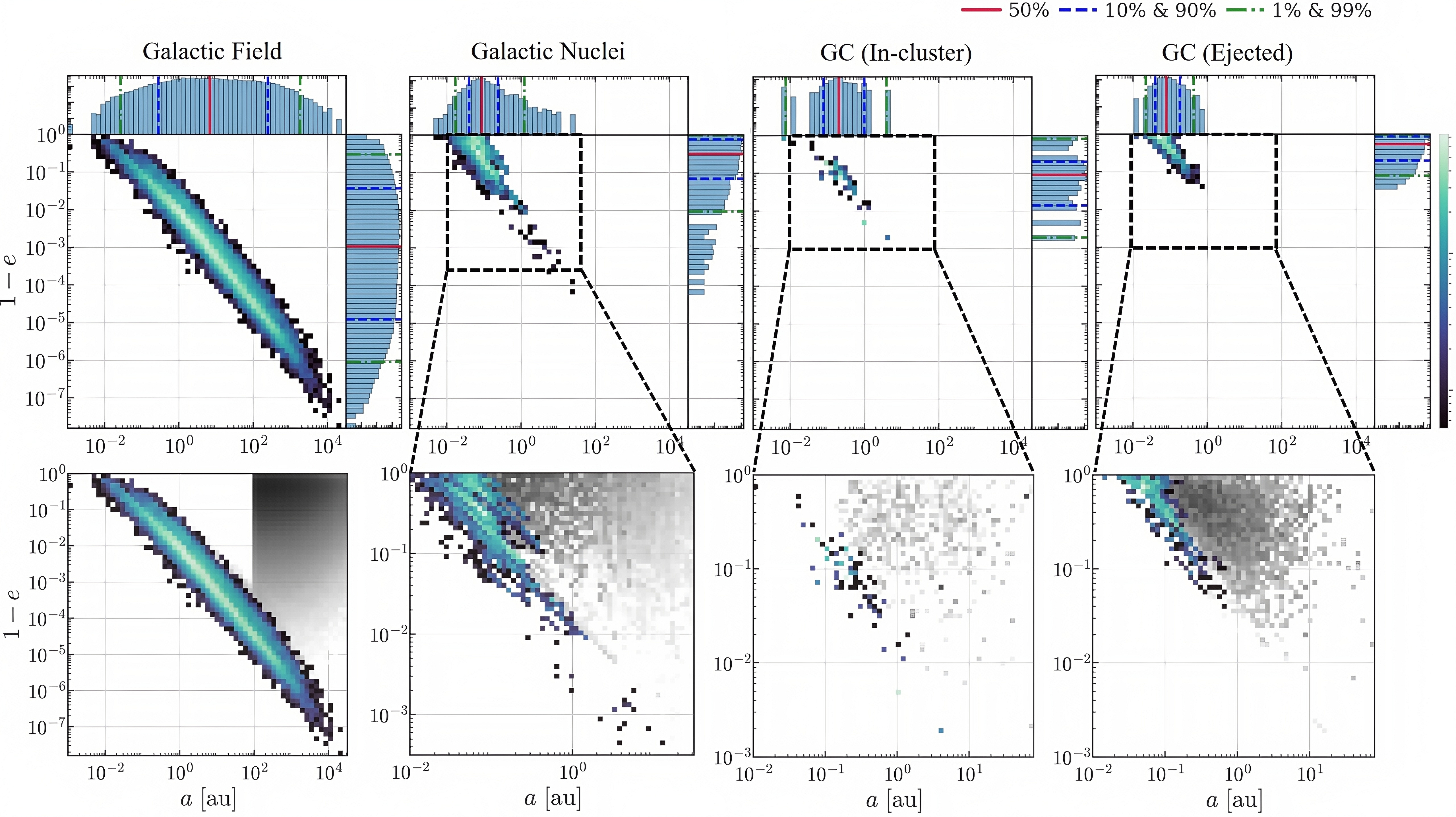}
    \caption{\textbf{Orbital parameter distributions of simulated Milky Way BBHs across different dynamical formation channels.} The four columns correspond to the Galactic Field, Galactic Nuclei, GC In-cluster, and GC Ejected BBH population, respectively. {\it Top panels} show the 2D probability density distributions in the $a$ and $1-e$ plane, for systems with $\mathrm{SNR} > 0.1$ for a \xzy{4-yr} LISA observation. The top and right sub-panels show the marginalized 1D probability density projections ($\mathrm{d}P/\mathrm{d}\log a$ and $\mathrm{d}P/\mathrm{d}\log(1-e)$). Vertical and horizontal lines indicate the 50\% (red solid), 10\% and 90\% (blue dashed), and 1\% and 99\% (green dash-dotted) percentiles of the distributions. {\it Bottom panels} show zoomed-in views of the regions containing the $\rm SNR>0.1$ populations, with the semi-transparent grey squares representing the underlying background population. Note that this figure assumes a local Milky Way snapshot. For populations at larger cosmological distances, the observable distribution would shift toward the upper-left region of the parameter space.}
    \label{fig:mw_bbh_distribution}
\end{figure*}
In this section, we examine the detailed population properties of the simulated BBH catalog. In particular, Figure~\ref{fig:mw_bbh_distribution} presents the orbital parameter distributions of BBHs formed through different channels, assuming a snapshot of the Milky Way is taken. The upper panels show the distribution of BBHs with $\rm SNR > 0.1$, in the $a$ -- $(1 - e)$ plane, with the color scale indicating the system number density. We also include the corresponding one-dimensional probability density projections along each axis (e.g., $\mathrm{d}P/\mathrm{d}\log a$ shown as histograms along the top axis), with the 1\%, 10\%, and 50\% percentiles marked by green dash-dotted, blue dashed, and red solid lines, respectively. The lower panels provide zoomed-in views of the parameter space for each population, with the remaining background population shown as semi-transparent grey points. Note that we show the distribution of BBHs with $\mathrm{SNR}>0.1$ to better illustrate their overall population properties. This should not be interpreted as the parameter distribution of detectable mHz BBHs, for which a typical detection threshold of $\mathrm{SNR}>8$ is adopted.

As shown in Figure~\ref{fig:mw_bbh_distribution}, different formation channels produce distinct orbital parameter distributions in the mHz GW source population. The Field (fly-by induced) channel is characterized by wider separations and more extreme eccentricities, with a median semi-major axis $a_{\rm mid} \sim 7\,\mathrm{au}$ and median eccentricity $e_{\rm mid} \sim 0.999$ (when they enter the mHz band with $\rm SNR>0.1$). In contrast, the Galactic Nucleus and globular cluster in-cluster populations are less extreme, with $a_{\rm mid} \sim 0.1$ and $0.2\,\mathrm{au}$, and $e_{\rm mid} \sim 0.7$ and $0.9$, respectively. The GC ejected population exhibits the smallest characteristic separations ($a_{\rm mid} \sim 0.09\,\mathrm{au}$) and the lowest median eccentricity ($e_{\rm mid} \sim 0.6$). As shown, the overall trends in the BBH orbital parameter distribution are consistent with the dynamical characteristics of their formation environments. Specifically, Field binaries are typically wider, such that only those driven to the extreme high-eccentricity tail can experience efficient GW energy loss and evolve towards the merger stage. In contrast, GN and GC in-cluster populations originate from intrinsically more compact progenitors (see the grey regions in Figure~\ref{fig:mw_bbh_distribution}), and frequent dynamical interactions in these dense environments (e.g., scattering and three-body encounters) drive a larger fraction of systems into the GW-dominated regime. In addition, the ejected GC population tends to retain lower eccentricities, while the high-eccentricity cluster binaries have shorter inspiral times and therefore tend to merge inside their host cluster before they can be dynamically ejected \citep[see, e.g.,][]{kremer2025compactobjectsglobularclusters}. Overall, these features provide guidance for targeted searches in LISA data and may help disentangle the contributions of different BBH formation channels in future observations.

We note that despite differences in orbital separations and eccentricities, mHz BBHs from different dynamical formation channels tend to occupy a narrow band in the $a$ -- $(1 - e)$ plane, corresponding to merger timescales $t_{\rm merger} \sim 10^4$--$10^{10}
\,\mathrm{yr}$ (see, e.g., Figure~\ref{fig:mock_mw} and  \ref{fig:mw_bbh_distribution}). This behavior arises from a selection effect similar to that discussed in Section~\ref{subsec:catalog}. In particular, systems are most likely to be detected by mHz GW detectors when they have a long lifetime ($t_{\rm merger}$), while a given remaining merger timescale approximately corresponds to a characteristic SNR level across the $a$–$(1 - e)$ parameter space. As a result, the observed snapshot population in the Milky Way is dominated by sources near the detection threshold ($\rm SNR \sim 5$ -- $100$), which occupies an elongated band in parameter space. 

We emphasize that these quantitative results apply specifically to the Milky Way population. For sources at larger cosmological distances, only binaries in later evolutionary stages can reach sufficiently high signal-to-noise ratios. Consequently, the observed population is biased toward smaller semi-major axes and lower eccentricities, corresponding to a shift toward the upper-left region of parameter space and toward shorter merger timescales. We further discuss extragalactic sources in Section~\ref{subsec:bbhcosmos}.

\begin{figure*}
    \centering
    \includegraphics[width=7in]{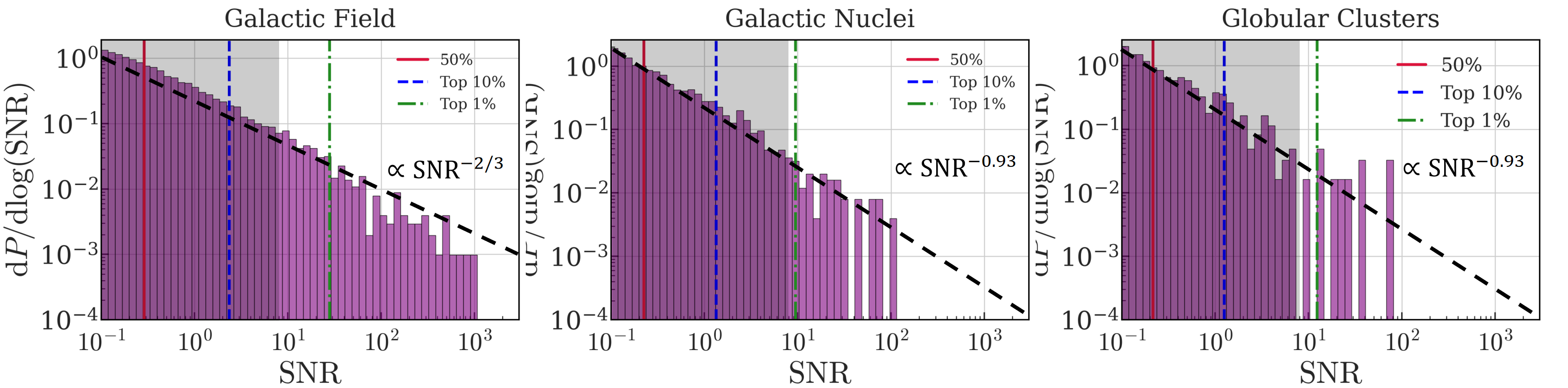}
    \caption{\xzy{\textbf{Expected number of dynamically-formed Milky Way BBHs in each SNR bin.} The histograms display the differential probability distribution $\mathrm{d}P/\mathrm{d}\log(\mathrm{SNR})$ for all the simulated systems with an expected $\mathrm{SNR} > 0.1$ in a 4-yr LISA observation. Vertical lines denote the 50\% (red solid), top 10\% (blue dashed), and top 1\% (green dash-dotted) percentiles of the population in SNR distribution. The shaded region marks systems below the nominal detectability threshold, $\mathrm{SNR}<8$, while the unshaded region corresponds to detectable sources. The black dashed lines illustrate the analytically derived power-law scaling, for comparison. Notably, the highly eccentric Field population follows a shallower slope ($\sim \mathrm{SNR}^{-2/3}$), whereas the GN and GC populations exhibit a steeper slope ($\sim \mathrm{SNR}^{-0.93}$). For a detailed analysis of these power-law dependencies, see Section~\ref{subsec:astro}.}}
    \label{fig:snrdistribution}
\end{figure*}

Figure~\ref{fig:snrdistribution} shows the SNR distribution of dynamically formed BBHs in the Milky Way. We compute the probability density in $\log \mathrm{SNR}$ for all sources with $\rm SNR > 0.1$ in the mHz band. The distributions for the Field, GN, and GC populations are shown separately, with the 1\%, 10\%, and 50\% percentiles indicated by vertical lines. As shown in the figure, the SNR distributions for different formation channels approximately follow power-law behavior. In particular, the GC and GN populations are consistent with a slope of $\sim -1$, while the Field population exhibits a slightly shallower slope ($\sim -2/3$). 

This power-law behavior can be understood analytically. In particular, randomly choosing an observation window $T_{\rm obs}$ during the long-term inspiral of a BBH system, the differential probability of the binary being observed emitting GW at a given SNR, $\mathrm{d}P/\mathrm{d}\mathrm{SNR}$, is proportional to the differential evolution time spent at that SNR, (i.e., $\mathrm{d}t_{\rm merger}/\mathrm{d}\mathrm{SNR}$, note that here we assume the binary evolves slowly, $T_{\rm obs}\ll t_{\rm merger}$). For BBHs with fixed masses, the merger timescale scales as $t_{\rm merger} \propto f_{\rm GW}^{-8/3}(1 - e)^{-1/2}$ (Equation~\ref{eq:lifetime}). Meanwhile, at a fixed source distance, the signal-to-noise ratio scales as ${\rm SNR}\propto S_n(f_{\rm GW})^{-1/2}f_{\rm GW}^{2/3}(1-e)^{3/4}$ (Equations~\ref{eq:snrnew} and \ref{eq:amplitude}). In the $\sim$1--3 mHz frequency range, the detector noise is dominated by the Galactic foreground, which approximately follows $S_n(f)\propto f^{-4.4}$ \citep[see, e.g.,][]{Klein+16}.

Combining these scalings, we get:
\begin{equation}
\begin{aligned}
t_{\rm merger} &\propto f_{\rm GW}^{-8/3}(1-e)^{-1/2} \ ,\\
{\rm SNR} &\propto f_{\rm GW}^{8.6/3}(1-e)^{3/4} \ .
\end{aligned}
\label{eq:snrandtmerger}
\end{equation}

Furthermore, when $e\rightarrow1$, the change of binary's pericenter distance is negligible during the GW-dominated orbital evolution, $dr_p=d[a(1-e)]\sim 0$ \citep[see eq. 5.8 in][]{Peters64}, which results in highly eccentric BBHs emitting GW with almost the same $f_{\rm GW}$ until $e$ drops to moderate values. In other words, the eccentricity evolution dominates the change in remaining merger time $t_{\rm merger}$ and observed SNR, and $f_{\rm GW}$ can be taken as approximately constant. In this case, Equation~(\ref{eq:snrandtmerger}) yields $t_{\rm merger} \propto (1-e)^{-1/2} \propto {\rm SNR}^{-2/3}$. On the other hand, when $e\rightarrow0$, the $(1-e)$ term in Equation~(\ref{eq:snrandtmerger}) has negligible influence on the system's evolution, and we approximately have $t_{\rm merger}\propto f_{\rm GW}^{-8/3}\propto{\rm SNR}^{-0.93}$. 

Finally, since the expected number of observed systems within a given logarithmic SNR bin is proportional to the fraction of the binary's lifetime spent in that state, we can analytically derive the functional forms shown in Figure~\ref{fig:snrdistribution}. Specifically, using the relation of $\mathrm{d}P/\mathrm{d}\log(\mathrm{SNR})={\rm SNR}\cdot dP/\mathrm{d}\mathrm{SNR}\propto {\rm SNR}\cdot\mathrm{d}t_{\rm merger}/\mathrm{d}\mathrm{SNR}$, and plug in the aforementioned power law relation between $t_{\rm merger}$ and SNR (i.e., Equation~(\ref{eq:snrandtmerger}), in the high e and low e limit):
\begin{equation}
\begin{aligned}
\frac{dP}{d\log(\mathrm{SNR})} &\propto {\rm SNR^{-2/3}}\,(\text{when}\, e\rightarrow1)\, ,\\
\frac{dP}{d\log(\mathrm{SNR})} &\propto {\rm SNR^{-0.93}}\,(\text{when}\, e\rightarrow0)\, .
\end{aligned}
\label{eq:snrdistribution_predict}
\end{equation}

We note that, Equation~(\ref{eq:snrdistribution_predict}) estimates the SNR distribution for a BBH population with fixed mass and distance to the observer, assuming its GW radiation is dominated by highly eccentric ($e\rightarrow1$) or quasi-circular sources ($e\rightarrow0$), respectively; In reality, the BBHs in the Milky Way have different distances and mass, and consist of both circular and eccentric systems. Therefore, the realistic slope in the SNR distribution may vary in the range of $\sim [-0.93,-0.67]$. Additionally, the power law relation we adopt here applies specifically for low-frequency source populations ($f_{\rm GW}\sim 1-3$\,mHz), while the GW emission of BBHs population at cosmological distances may be dominated by high frequency components, thus having a different power law relation. However, Equation~(\ref{eq:snrdistribution_predict}) still matches well and further verifies the simulation results in Figure~\ref{fig:snrdistribution}. In particular, the eccentric Field BBH population exhibits a shallower slope $\sim -0.67$, while the GN and GC populations, characterized by less extreme eccentricities, show steeper slopes closer to $\sim -0.93$.

\subsection{BBHs at Cosmological Distances}
\label{subsec:bbhcosmos}

\begin{figure}
    \centering
    \includegraphics[width=3.4in]{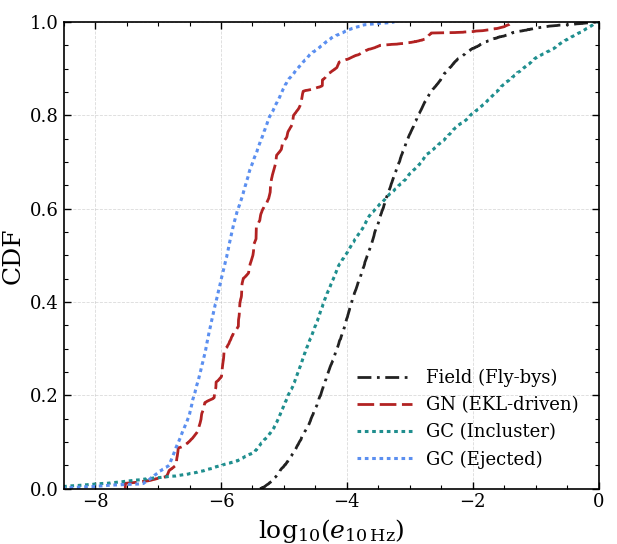}
    \caption{\textbf{Eccentricity distribution of dynamically-formed BBHs in the LIGO band.} The plot compares the cumulative distribution functions (CDFs) of eccentricities at $f_{\rm GW} = 10\,\mathrm{Hz}$, for BBH mergers formed via Galactic Field fly-bys (black dot-dashed line), Galactic Nucleus EKL-driven mergers (red dashed line), and two Globular Cluster populations (in-cluster mergers as the teal dotted line, and ejected mergers as the blue dotted line). The GC distributions are adapted from \citet{Martinez+20}. The in-cluster GC curve shown here combines contributions from the KL and non-KL triple channels, few-body and single-single capture channels, and the in-cluster binary channel, following their fig.~4. \xzy{This figure provides a unified comparison of the eccentricity distributions of dynamically formed BBHs at the final merger stage, combining previously published GC results with new predictions for the eccentricity distribution of Galactic Field fly-by and Galactic Nucleus EKL-induced merger channels.}}
    \label{fig:ecc_cdf}
\end{figure}

We further explore the population of eccentric BBHs at cosmological distances, for both the LIGO and LISA detection. Specifically, we adopt the same population model as in Section~\ref{subsec:catalog}, evolve the Field and GN populations forward, and compute their eccentricities when entering the LIGO band at $f_{\rm GW} = 10\,\mathrm{Hz}$. The resulting distributions are shown in Figure~\ref{fig:ecc_cdf}. For comparison, we also include the eccentricity distributions of in-cluster and ejected GC populations from fig.~4 of \citet[][]{Martinez+20}. Here the in-cluster GC curve combines contributions from the triple, capture, and in-cluster binary channels in their results. For consistency, we adapt our peak GW frequency definition, $f_{\rm GW}=2f_{\rm orb}(1-e)^{-1.5}$, to their eq.~16, $f_{\rm GW}=2f_{\rm orb}(1-e)^{-1.5}(1+e)^{-0.3046}$ in Figure~\ref{fig:ecc_cdf}. These two definitions differ by a factor of order unity in the high-e regime, and both reduce to $f_{\rm GW}=2f_{\rm orb}$ in the low-e limit.

\begin{figure*}
    \centering
    \includegraphics[width=7in]{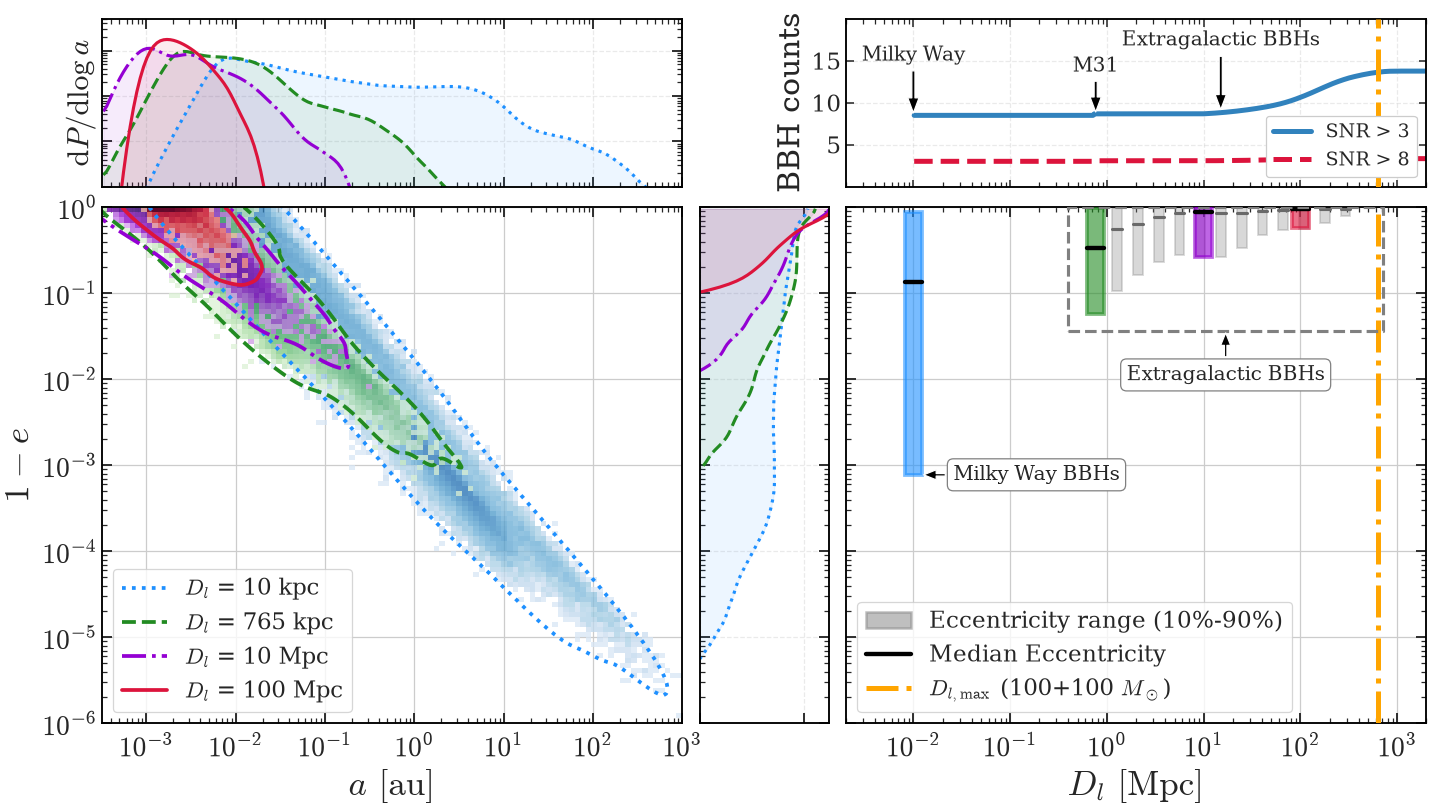}
    \caption{\textbf{Orbital properties and number of mHz eccentric BBHs at different luminosity distances.} \xzy{All panels assume a 4-year LISA observation and a SNR threshold of $8$.} The {\it bottom-left panel} shows the 2D probability density distribution of the detectable dynamically-formed BBH population in the $a$ -- $(1-e)$ parameter space. Overlaid contours indicate the $99\%$ percentiles for populations placed at different luminosity distances: $10\,\mathrm{kpc}$ (blue dotted), $765\,\mathrm{kpc}$ (green dashed), $10\,\mathrm{Mpc}$ (purple dot-dashed), and $100\,\mathrm{Mpc}$ (red solid). The {\it top-left} and {\it center vertical} panels display the corresponding 1D marginal distributions for the semi-major axis ($\mathrm{d}P/\mathrm{d}\log a$) and eccentricity ($\mathrm{d}P/\mathrm{d}\log(1-e)$), respectively. The {\it top-right panel} illustrates the cumulative number of expected BBHs as a function of distance, starting from the Milky Way local contribution to the extragalactic population. The {\it bottom-right panel} presents the 1-D eccentricity distribution (10\% to 90\% percentiles, with black median ticks) at different $D_l$, with highlighted color bars directly matching the populations in the left panels. The vertical orange dot-dashed line across the right panels marks the maximum detectable distance ($D_{l,\rm max} \sim 1\,\mathrm{Gpc}$) for a fiducial binary mass of $100+100\,M_\odot$. As distance increases, the observable BBH population systematically shifts toward tighter and more circular orbits due to the selection effect.}
    \label{fig:orb-dl}
\end{figure*}
As shown in Figure~\ref{fig:ecc_cdf}, the merger eccentricities from different formation channels span a wide range, with an overall trend similar to that in the mHz band. In particular, globular cluster in-cluster mergers and field fly-by induced mergers tend to retain higher eccentricities, which are potentially detectable by LIGO when entering the 10 Hz band. In contrast, although GN and GC ejected populations may have high eccentricities at formation (see Figure~\ref{fig:mw_bbh_distribution}), their residual eccentricities in the LIGO band are typically very small ($\lesssim 10^{-2}$), making them nearly indistinguishable from mergers formed via isolated binary evolution.

\xzy{This channel dependence has important implications for interpreting BBH mergers in the LVK data. For example, recent studies have explored using eccentric detections, or their null detection, to constrain the contribution of dynamically-formed BBHs in the observed merger population \citep[e.g.,][]{Zevin_2021}. Our results suggest that such constraints should be interpreted on a channel-by-channel basis. In particular, many dynamically-formed binaries are born with high eccentricity, but their eccentricities are efficiently damped through GW emission long before entering the LVK band. As a result, the absence of eccentric detections primarily constrains those channels that retain measurable eccentricities, while many other dynamical channels, such as GN EKL-driven mergers and GC-ejected binaries, may still contribute substantially to the LVK merger rate despite appearing nearly circular in detection.}

On the other hand, as discussed in Section~\ref{subsec:astro}, the detectability of wide, highly eccentric sources in the mHz band is limited to the local Universe. Nevertheless, the general population of dynamically-formed BBHs can still be observed at cosmological distances, provided that they are approaching the final evolution stages. For example, a $100+100\,M_\odot$ BBH can be detected up to $\sim 1\,\mathrm{Gpc}$ for a 4-year LISA observation (see the right panel of Figure~\ref{fig:orb-dl}). As the distance increases, the detectable population gradually shifts toward less eccentric, more compact, and shorter-lived systems, as they have a larger signal-to-noise ratio.

This trend is reflected in Figure~\ref{fig:orb-dl}. In particular, we adopt the same population model as in Section~\ref{subsec:catalog}, place sources at different distances, and compute their orbital parameter distributions \xzy{for systems with signal-to-noise ratio $\mathrm{SNR} > 8$ over a 4-year LISA observation.} The {\it bottom-left panel} of Figure~\ref{fig:orb-dl} shows the population in the $a$ -- $(1-e)$ parameter space at distances of 10 kpc, 765 kpc, 10 Mpc, and 100 Mpc (blue, green, purple, and red contours, respectively), along with their corresponding 1D marginal distributions on the top and right axes. The {\it bottom-right panel} further shows the distribution of eccentricity (10\%–90\% percentiles and median) as a function of distance. As illustrated, in the local Universe (e.g., the Milky Way at 10 kpc, or M31 at 765 kpc), a large number of wide, highly eccentric systems are detectable, with median eccentricities around $e \sim 0.9$ and a significant fraction of BBHs reaching extreme eccentricity values ($e \gtrsim 0.999$). As the distance increases, the detectable parameter space shrinks toward the upper-left region, and only less eccentric and more compact systems remain observable. For example, at distances of $\sim 10$ Mpc, the median eccentricity drops to $e \lesssim 0.1$, and the semi-major axis is typically below $10^{-2}$ AU, indicating a transition to typical mHz quasi-circular inspirals.

In addition to the orbital properties, the {\it top-right panel} of Figure~\ref{fig:orb-dl} demonstrates the estimated cumulative number of detectable BBHs as a function of distance. Specifically, we adopt the Milky Way simulation's result ($\sim 9$ sources with $\rm SNR>3$ at $D_l\sim 10^{-2}$~Mpc), and consider the contribution from nearby galaxies, such as M31. Assuming M31 hosts a BBH population similar to the Milky Way but scaled by its stellar mass (approximately twice as large), we find the expected detectable source number is $\sim 0.2$ (due to the larger distance and weaker signal strength), which indicates the nearby galaxies have a limited contribution to the eccentric BBHs detection. 

We then incorporate the cosmological contribution by assuming that extragalactic BBHs follow similar orbital parameter distributions and population properties as in our Milky Way model. The corresponding merger rates are taken directly from our simulation, yielding $\sim 3$, $2$, and $4\times10^{-7}\,\mathrm{yr}^{-1}$ per Milky-Way-like galaxy for the Field, GN, and GC channels, respectively. These rates translate to volumetric merger rates of $\sim 3$, $2$, and $4~\mathrm{Gpc}^{-3}\,\mathrm{yr}^{-1}$, assuming a galaxy number density of $0.01~\mathrm{Mpc}^{-3}$ \citep{Conselice_2005, van_Dokkum_2013}. For each distance, we compute the average detectable lifetime of BBHs \xzy{(for a 4-yr LISA observation)} and multiply it by the merger rate within the corresponding comoving volume shell to estimate the expected number of observable sources.

As shown in the cumulative curve in the {\it top-right panel}, BBHs at cosmological distances (e.g., $10$–$1000$ Mpc) provide a significant background population, yielding $\sim 6(150)$ sources with $\mathrm{SNR} > 3(1)$. However, the number of detectable sources decreases rapidly with an increasing SNR threshold. For example, at $\mathrm{SNR} = 8$, the expected number of extragalactic sources drops to $\lesssim 1$. This indicates that extragalactic sources are typically faint and contribute primarily at low SNR, with a steeper decline toward higher SNR compared to nearby highly eccentric sources (see Table~\ref{tab:snr_stats}).

We note that our calculation assumes that BBHs at cosmological distances follow the same parameter distributions as those in our local dynamical simulations, which may introduce biases. In addition, our model does not account for all the potential dynamically-formed BBH populations. Nevertheless, our results illustrate a robust picture for the detection of eccentric BBHs in the mHz band. Specifically, the observed population is dominated by two regimes: very local sources in the Milky Way, which are wide and highly eccentric and are detectable due to their proximity and long lifetimes; and extragalactic populations at $\sim 100$–$1000$~Mpc, which are more compact and nearly circular and are detectable due to the large detection volume, but typically have lower signal-to-noise ratios.

\section{Eccentric BBHs in the LISA Global Fit}\label{sec:globalfit}

\begin{figure}[htbp]\centering\includegraphics[width=3.3in]{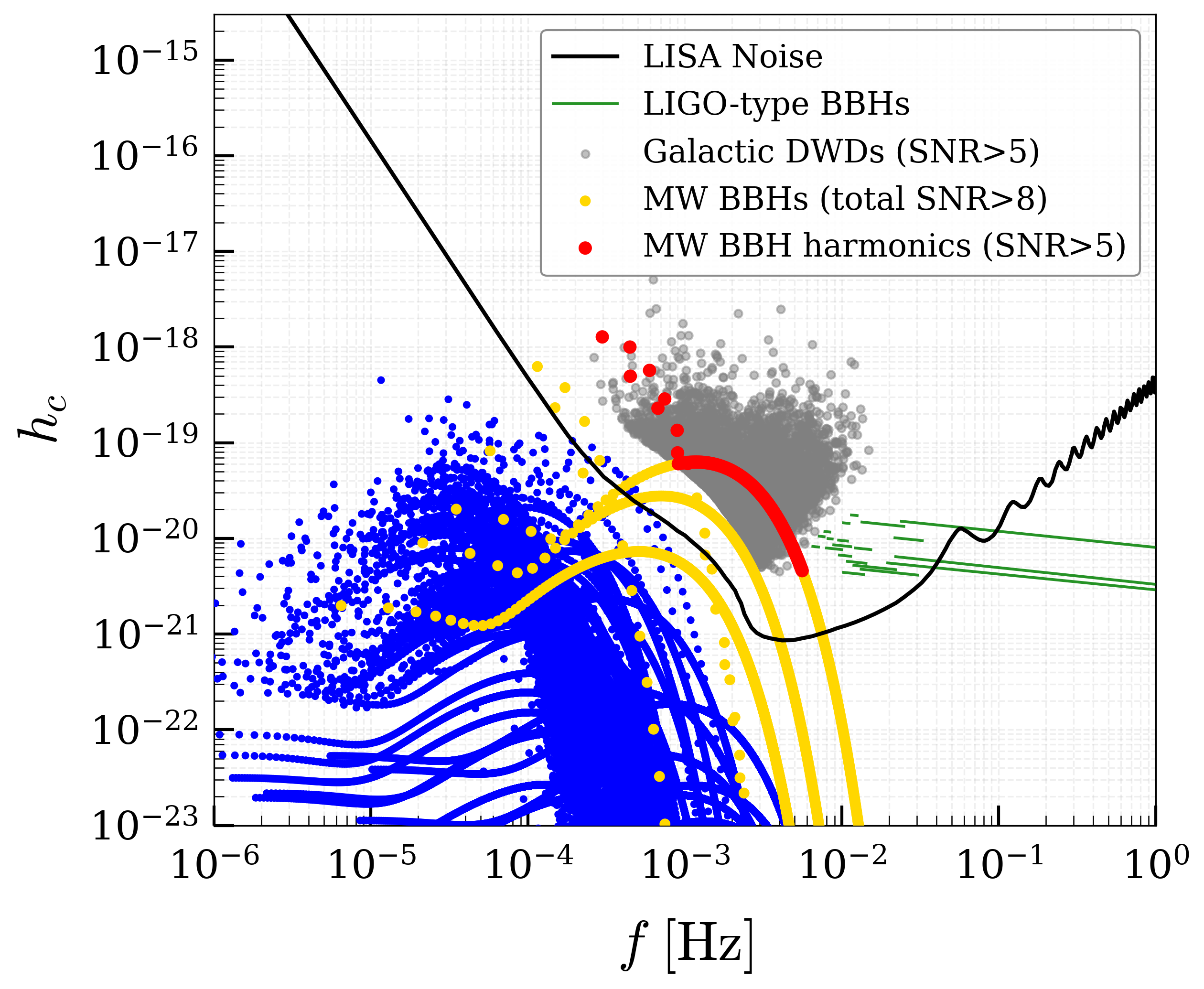}\caption{\textbf{Individual GW harmonics of dynamically-formed BBHs in one realization of the Milky Way.} The plot shows the characteristic strain for dynamically-formed BBHs in a single realization of the Milky Way ({\it individual harmonic representation} $h_{c,n}$, see Equation~\ref{eq:hcn,0}). The red markers highlight GW harmonics that have an individual SNR $> 5$. The blue points represent the background harmonics from all GW sources with a total SNR $\geq 0.1$ for a 4-yr LISA observation, and the yellow points represent BBHs with total SNR (summing over all harmonics) greater than 8, but individual harmonic SNR smaller than 5. For comparison, we also plotted the Galactic double white dwarf population \citep[grey point, see, e.g., ][]{lamberts18} and extragalactic circular BBHs \citep[green curves, see, e.g., ][]{amaro+22}. }\label{fig:hcscatter}\end{figure}

As previously discussed, dynamically-formed BBHs can constitute a significant population in the mHz GW band. Their signals are characterized by multiple-harmonic structures that span a broad frequency range. This distinct feature presents both opportunities and challenges for practical signal processing. In this section, we will focus on the confusion that their multiple harmonic structures may cause in the LISA global fit, as well as the corresponding stochastic background.
\citep[][]{Xuan+21,Chen20fakebbh}

Specifically, the current global fit often relies on quasi-circular templates to search for Galactic binaries. However, highly eccentric sources possess multiple harmonics, and these harmonics often appear as slowly-evolving, near-monochromatic components. Without introducing specific search methods to separate them, the signal of a single eccentric BBH could be misidentified as multiple circular signals at different harmonics. For example, Figure~\ref{fig:hcscatter} shows a mock realization of dynamically formed BBHs in the Milky Way (which is the same mock population as shown in Figure~\ref{fig:mock_mw}). We plot the characteristic strain for all the harmonics from MW BBHs with $e>0.1$ \xzy{(assuming a 4-yr LISA observation, see Equation~\ref{eq:hcmixed})}, as shown by the blue points. In addition, we highlight the harmonics with an individual SNR greater than 5 in red, and harmonics with an individual $\rm SNR<5$ but correspond to a BBH source with a combined $\rm SNR>8$ in yellow. For comparison, we also plotted the Galactic double white dwarf population (grey points) and extragalactic, LIGO-type BBHs \citep[green curves, see, e.g., ][]{robson18}.

As shown by Figure~\ref{fig:hcscatter}, while the majority of eccentric MW BBHs are not detectable, a subset of systems can exhibit multiple harmonics that exceed the detection threshold. Due to their extreme eccentricities, once a system crosses the detection threshold, it typically produces a dense sequence of resolvable harmonics, which inevitably introduces the risk of signal overlap and confusion, particularly if they are inadvertently analyzed using standard quasi-circular templates. In addition, some BBHs (in yellow) do not have harmonics lying beyond the noise curve, while they still have a total $\rm SNR>8$. This is because here we are plotting the energy of individual harmonics. For highly eccentric systems, the density of harmonics is extremely large, which, when summed together, yield a significant SNR. Therefore, if we adopt the {\it smooth spectrum representation} as explained in Appendix~\ref{app:plothighe}, these systems will have a spectrum curve beyond the noise curve.

\begin{figure}[htbp]\centering\includegraphics[width=3.2in]{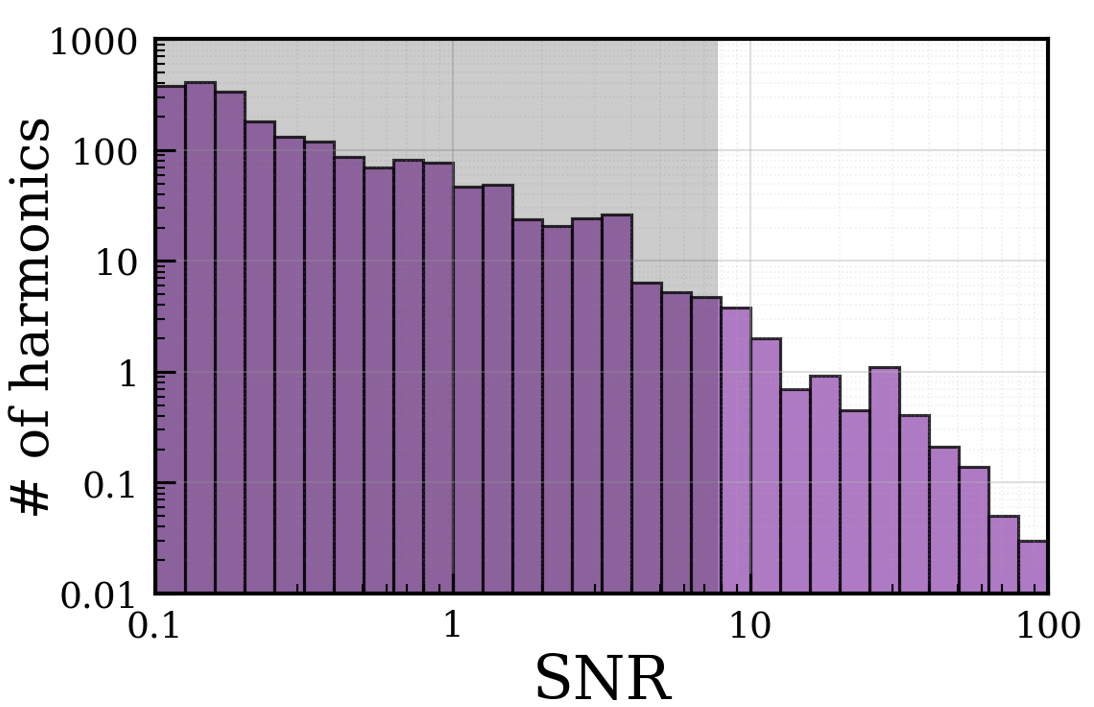}\caption{\xzy{\textbf{Distribution of individual GW harmonics in each SNR bin, for BBHs in the Milky Way.} The histogram displays the average number of individual GW harmonics per Milky Way realization as a function of their single-harmonic SNR. The purple bars represent the expected count of harmonics within each log-uniform SNR bin, derived by averaging over 100 realizations of the Galactic population. The shaded region marks harmonics below the nominal detectability threshold, $\mathrm{SNR}<8$, while the unshaded region corresponds to detectable harmonics. The distribution highlights the relative abundance of sub-threshold harmonics versus individually resolvable harmonics for a 4-yr LISA observation.}}\label{fig:SNR_HARM}\end{figure}

Figure~\ref{fig:SNR_HARM} further quantifies the expected number of individual detectable GW harmonics from bursting BBHs across different SNR bins, averaged over 100 mock realizations of the Milky Way. As demonstrated, dynamically formed BBHs can contribute an average of $\sim 216,\, 60,\, 10,\, \text{and } 3$ individual harmonics with $\rm SNR > 1,\, 3,\, 8,\, \text{and } 20$, respectively, to the LISA data stream, which can significantly complicate source extraction. Notably, our results show that these counts are characterized by large standard deviations, typically comparable to or even exceeding the aforementioned average numbers. This massive variance arises from the highly non-Gaussian nature of the harmonic distribution. In a typical realization, the entire population of detectable harmonics originates from only $1-3$ high-SNR sources. However, once such a source is present with high eccentricity, it simultaneously produces a substantial cluster of harmonics with intrinsically correlated frequencies and SNRs.

To further understand the extent of this overlap and confusion, we derive the parameter bias introduced when an individual harmonic of an eccentric BBH is mistakenly identified as the dominant $2f_{\mathrm{orb}}$ mode of a circular binary. In particular, for an eccentric system with a true chirp mass $\mathcal{M}_{c,0}$ and orbital eccentricity $e$, the orbital frequency $f_{\mathrm{orb}}$ evolution can be estimated using \citep{Peters64}:

\begin{equation}\dot{f}_{\mathrm{orb}} = \frac{96}{5}\pi^{8/3} \mathcal{M}_{c,0}^{5/3} f_{\mathrm{orb}}^{11/3} F(e)  2^{8/3}\, ,
\end{equation}
where $F(e) = (1 + \frac{73}{24}e^2 + \frac{37}{96}e^4) / (1 - e^2)^{7/2}$ is the eccentricity enhancement factor. The factor of $2^{8/3}$ arises because here we use the orbital frequency $f_{\rm orb}$ instead of the gravitational wave frequency.

Since the frequency of the $n$-th harmonic is $f_n = n f_{\mathrm{orb}}$, its time derivative is $\dot{f}_n = n \dot{f}_{\mathrm{orb}}$. Rearranging the terms to isolate $f_n$, we obtain:
\begin{equation}\dot{f}_n = \frac{96}{5}\pi^{8/3} \mathcal{M}_{c,0}^{5/3} \left[ F(e)  2^{8/3}  n^{-8/3} \right] f_n^{11/3}\,.
\end{equation}

If a data analysis pipeline fits this individual spectral track using the standard circular chirp formula, $\dot{f} = \frac{96}{5}\pi^{8/3}(\mathcal{M}'_c)^{5/3} f^{11/3}$, the inferred apparent chirp mass $\mathcal{M}_c'$ will be biased compared with the true chirp mass of the eccentric source:
\begin{equation}\label{eq:apparent_mc}\mathcal{M}'_c = \mathcal{M}_{c,0}  F(e)^{3/5} \left(\frac{2}{n}\right)^{8/5}\,.
\end{equation}

In addition, the gravitational wave radiation of highly eccentric systems is strongly localized at periapsis, which produces a broadband signal peaked at the harmonic $n_{\mathrm{peak}} \sim 2(1 - e)^{-3/2}$ (see Equation~\ref{eq:peakf}). Substituting this peak harmonic estimate into Equation~(\ref{eq:apparent_mc}) yields the analytic mass bias near the peak radiation frequency:
\begin{equation}\left.\frac{\mathcal{M}'_c}{\mathcal{M}_{c,0}}\right|_{n=n_{\rm peak}} = \frac{(1 - e)^{3/10}}{(1 + e)^{21/10}} \left( 1 + \frac{73}{24}e^2 + \frac{37}{96}e^4 \right)^{3/5} \ . \end{equation}

\begin{figure}[htbp]\centering\includegraphics[width=3.4in]{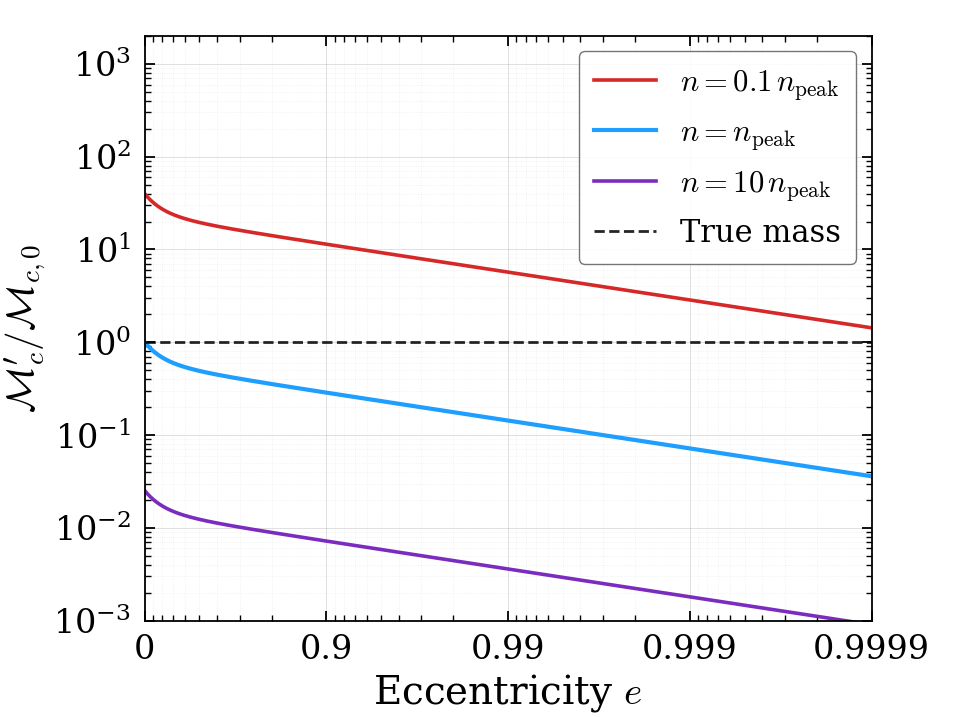}\caption{\textbf{Apparent chirp mass bias for highly eccentric sources.} The curve demonstrates the ratio of the inferred apparent chirp mass $\mathcal{M}_c'$ to the true chirp mass $\mathcal{M}_{c,0}$ when the peak radiating harmonic of an eccentric binary is mistakenly fitted with a quasi-circular waveform template. As the eccentricity approaches unity, the apparent mass significantly underestimates the true system mass.}\label{fig:fakemc}\end{figure}

This analytic mapping is visualized in Figure~\ref{fig:fakemc}. Because of the strong $(1-e)^{0.3}$ dependence in the numerator, as $e \to 1$, this ratio approaches zero. Consequently, a highly eccentric stellar-mass BBH radiating near its peak frequency would mimic a circular binary with a much smaller chirp mass. This degeneracy highlights the danger of mischaracterizing individual harmonics if they are not jointly analyzed using a fully eccentric template framework.

We note that the bias derived here is only a rough estimate based on matching the first frequency derivative, $\dot{f}$, rather than a full waveform fit. In principle, this degeneracy could be broken by considering higher-order frequency evolution. For circular binaries, the second derivative $\ddot{f}$ is uniquely determined by $\dot{f}$, whereas for eccentric harmonics the additional dependence on harmonic number $n$ and eccentricity $e$ breaks this relation. A measurement of $\ddot{f}$ could therefore, in principle, distinguish between the two cases. In practice, however, the local BBH population consists of long-lived systems with merger timescales typically exceeding $\sim 10^6$ yr (see Equation~\ref{eq:lifetime}), making them effectively non-chirping over a 4-10 yr observation window. For such sources, the observational sensitivity to $\ddot{f}$ is orders of magnitude below the intrinsic $\ddot{f}$ of the dominant harmonics, so the signals are effectively characterized by $\dot{f}$ alone, or are nearly monochromatic. In this regime, $\ddot{f}$ cannot realistically break the degeneracy, and the apparent chirp-mass bias derived above therefore remains relevant \citep[see also][]{Xuan+21,Chen20fakebbh}.

\begin{figure}\centering\includegraphics[width=3.2in]{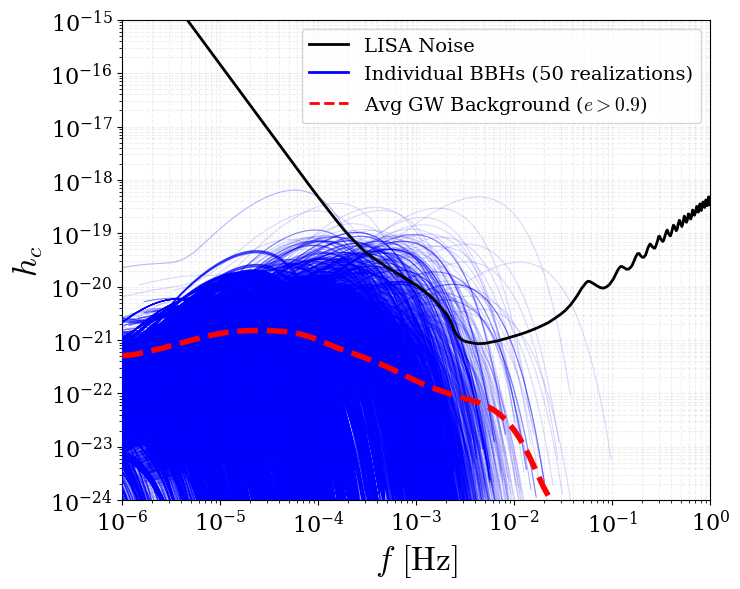}\caption{\textbf{Characteristic strain spectrum and collective GW background of highly eccentric BBHs in the Milky Way.} The plot shows the characteristic strain for BBH sources with $e > 0.9$ in the Milky Way ({\it smoothed spectrum representation} $h_{c,\rm env}$, see Equation~\ref{eq:hc_highe0}). The thin blue curves represent the characteristic strains of dynamically formed BBHs, compiled from 50 realizations of our mock Galactic catalog, \xzy{assuming a 4-year LISA observation.} The thick red dashed curve illustrates the average stochastic GW background produced by this highly eccentric subpopulation in a single MW realization, calculated following eq.19 in \citet{Xuan24bkg}. For reference, the solid black curve denotes the nominal LISA instrumental noise.}\label{fig:hc}\end{figure}

Additionally, the collective GW background of highly eccentric BBHs can also impact the LISA global fit. For example, Figure~\ref{fig:hc} displays the characteristic strain envelope and the stochastic GW background from the simulated highly eccentric BBHs ($e > 0.9$) in the Milky Way. As shown, the collective background from the Milky Way BBH population alone approaches the nominal LISA noise curve. In fact, if we additionally account for extragalactic high-eccentricity sources, this stochastic background will exceed the LISA instrumental noise within its most sensitive mHz band \citep[see][for more details]{Xuan24bkg}. This implies that if high-eccentricity GW bursts are not appropriately modeled and subtracted, they can become a substantial astrophysical foreground, which could interfere with the detection and parameter estimation of other mHz GW sources. The specific implementation of these eccentric sources within the comprehensive LISA global fit pipeline is left for future work.

\section{Waveform Generation and Validation}
\label{sec:waveform}
\begin{figure*}
    \centering
    \includegraphics[width=7in]{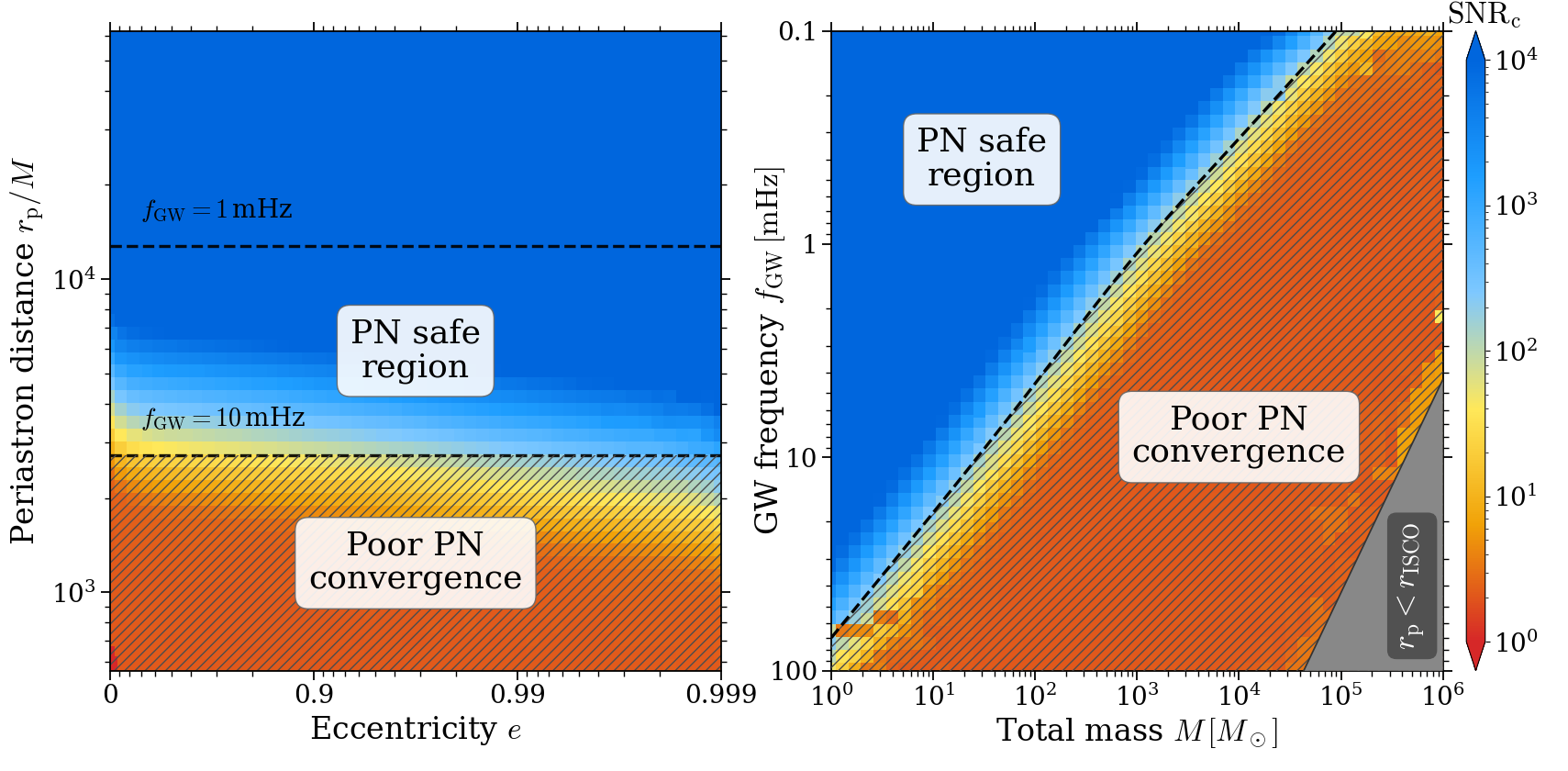}
    \caption{\textbf{Parameter spaces for post-Newtonian waveforms to converge well in eccentric GW data analysis.} The colormaps show the critical signal-to-noise ratio ($\mathrm{SNR}_c$) required to distinguish between waveforms generated using 2PN and 3PN expansions over a 1-year observation, computed numerically using waveform fitting methods (Equation~\ref{eq:snr_c}). The red hatched regions (``Poor PN convergence'') indicate parameter spaces where the PN expansion may break down ($\mathrm{SNR}_c \lesssim 100$), making it inadequate for data analysis \cite[see also][note that their fig.1 compares PN with PM expansion, whereas we focus on the {\it intrinsic} convergence of PN expansion.]{Khalil_2022}. {\it Left Panel:} $\mathrm{SNR}_c$ as a function of periastron distance $r_{\rm p}$ and eccentricity $e$ for a fixed binary mass of $20+25\,M_\odot$. The horizontal dashed lines mark constant GW peak frequencies of $1\,\mathrm{mHz}$ and $10\,\mathrm{mHz}$. As shown, PN waveforms are highly accurate for stellar-mass systems with high eccentricities, provided $f_{\rm GW} \lesssim 10\,\mathrm{mHz}$. {\it Right Panel:} $\mathrm{SNR}_c$ as a function of $f_{\rm GW}$ and total mass $M$, assuming an equal-mass binary with $e = 0.9$. The solid gray region is excluded because the periastron distance falls below the innermost stable circular orbit ($r_{\rm p} < r_{\rm ISCO}$, in which $r_{\rm ISCO}=6M$ for a Schwarzschild BH). It demonstrates that as total mass increases, PN waveforms can only converge at lower frequencies for mHz GW data analysis.}
    \label{fig:pn_accuracy}
\end{figure*}

Accurate waveform modeling is essential for the detection and characterization of eccentric BBHs in GW data analysis. However, the validity of existing GW waveform models for highly eccentric systems remains poorly understood. In particular, some analytical models used in GW data analysis rely on small-eccentricity expansions \citep[e.g.,][]{Buonanno_2009, Moore_2019, LALsuite}, which may break down in the high-eccentricity regime. Furthermore, as the eccentricity increases, changes between the pericenter and apocenter introduce significant dynamical scale differences, posing challenges for constructing surrogate models calibrated to numerical relativity simulations.

Nevertheless, for eccentric, stellar-mass mHz sources, post-Newtonian waveforms based on the quasi-Keplerian parametrization can still be effective. This is because, first, the conservative orbital dynamics in quasi-Keplerian parametrization are expressed in closed form, which does not require a Taylor expansion in eccentricity and can mathematically treat arbitrarily large $e$. Second, for these systems, the pericenter distance can remain sufficiently large for the post-Newtonian expansion to be valid; in particular, it can be related to the peak gravitational-wave frequency as:
\begin{equation}
r_p = \left( \frac{M}{\pi^2 f_{\rm GW}^2} \right)^{1/3},
\label{eq:rpfgw}
\end{equation}
which implies
\begin{equation}
\begin{aligned}
\frac{r_p}{r_s} &= \frac{1}{2 (\pi M f_{\rm GW})^{2/3}} \\
&\sim 1.1\times10^4\left(\frac{M}{20\,\rm M_{\odot}}\right)^{-2/3}
\left(\frac{f_{\rm GW}}{1\,\rm mHz}\right)^{-2/3}\, ,
\end{aligned}
\label{eq:rprs}
\end{equation}
where $r_s = 2M$ is the Schwarzschild radius corresponding to the total mass of the binary. Equation~(\ref{eq:rprs}) can also be related to the binary's pericenter velocity via $\left( v_p / c \right)^2 \simeq \frac{1}{2}(r_p/r_s)^{-1}(1+e)$.

Therefore, even for extremely eccentric BBHs, the dynamics near pericenter remain well within the weak-field and low-velocity regime (i.e., $r_p/r_s \sim 10^{4}$ and $v_p/c \sim 10^{-2}$), provided the system lies in the stellar-mass range and emits gravitational waves in the mHz band. This suggests that the PN expansion converges well and can be appropriate for modeling eccentric inspiral signals in the LISA band.

We can further quantify the convergence of the PN expansion by computing the overlap between waveforms of different PN orders. For example, setting $h_1$ as the 2PN waveform model with GW source parameters $\boldsymbol{\lambda}_1$, and $h_2$ as the 3PN waveform model with GW source parameters $\boldsymbol{\lambda}_2$, the normalized overlap between $h_1(\boldsymbol{\lambda}_1)$ and $h_2(\boldsymbol{\lambda}_2)$ is defined as \citep[see, e.g., section IV of][]{Afle+18}: 
\begin{equation}
\mathcal{O}(h_{1}(\boldsymbol{\lambda}_1), h_{2}(\boldsymbol{\lambda}_2))=\frac{\langle h_{1}(\boldsymbol{\lambda}_1)| h_{2}(\boldsymbol{\lambda}_2)\rangle}{\sqrt{\langle h_{1}(\boldsymbol{\lambda}_1)| h_{1}(\boldsymbol{\lambda}_1)\rangle\langle h_{2}(\boldsymbol{\lambda}_2)| h_{2}(\boldsymbol{\lambda}_2)\rangle}} \ ,
\label{eq:overlap}
\end{equation}
in which $\langle{h}_{1}| {h}_{2}\rangle$ represents the noise-weighted inner product between $h_{1}$ and $h_{2}$: 
\begin{equation}
\left\langle h_{1} \mid h_{2}\right\rangle=2 \int_{0}^{\infty} \frac{\tilde{h}_{1}(f) \tilde{h}_{2}^{*}(f)+\tilde{h}_{1}^{*}(f) \tilde{h}_{2}(f)}{S_{\mathrm{n}}(f)} \mathrm{d} f \ ,
\label{eq:innerproduct}
\end{equation}
where $\tilde{h}_j$ (with $j=1,2$) stands for the Fourier transformation
of the waveform, and the star stands for the complex conjugate.

The overlap in Equation~(\ref{eq:overlap}) reflects the similarity between waveforms $h_1(\boldsymbol{\lambda}_1)$ and $h_2(\boldsymbol{\lambda}_2)$. For example, the absolute value of $\mathcal{O}(h_{1}, h_{2})$ varies from $0$ to $1$, and a perfect match between two waveforms would give $\mathcal{O}(h_{1}, h_{2}) = 1$. 

Furthermore, the fitting factor (FF) between these two waveform families, $h_{1}$ and $h_{2}$, is defined as the maximized overlap over the signal's arrival time $t_{0}$, phase $\Phi_{0}$, and all the template parameters $\boldsymbol{\lambda}$:
\begin{equation}
\rm {F F}=\max_{\boldsymbol{\lambda}_2, t_0, \Phi_0}  \{\mathcal{O}(h_{1}(\boldsymbol{\lambda}_1), h_{2}(\boldsymbol{\lambda}_2))\} \ .
\label{eq:FF}
\end{equation}

In reality, FF exceeding a given threshold, as set by the signal-to-noise ratio, means the waveform models $h_1$ and $h_2$ are indistinguishable (i.e., equally good) for data analysis
\citep{lindblom08,2017PhRvD..95j4004C,2002ApJ...575.1030T,Cutler+07}: 
\begin{equation}
	{\rm FF}> 1-(D-1)/(2~\rm{SNR}^2)\ ,
	\label{eq:criteriaFF}
\end{equation}
in which $D$ represents the dimension of parameter space where the data analysis is carried out ($\sim10$ for LISA-band sources, depending on the template). 

However, since our focus is on the {\it intrinsic} convergence of PN expansion, we will fix the template parameters ($\boldsymbol{\lambda}_1=\boldsymbol{\lambda}_2=\boldsymbol{\lambda}$) in the following comparison, computing the overlap using Equation~(\ref{eq:overlap}), and directly assessing it against the right side of Equation~(\ref{eq:criteriaFF}):
\begin{equation}
\mathcal{O}(h_{1}(\boldsymbol{\lambda}), h_{2}(\boldsymbol{\lambda}))=\left.
\frac{\langle h_{1}| h_{2}\rangle}{\sqrt{\langle h_{1}| h_{1}\rangle\langle h_{2}| h_{2}\rangle}}
\right|_{\boldsymbol{\lambda}}
> 1-\frac{D-1}{2~\rm{SNR}^2}\ .
\label{eq:criteria_full}
\end{equation}
Because the fixed-parameter overlap is always a lower bound on the fitting factor, $\mathrm{FF}\geq\mathcal{O}(h_1(\boldsymbol{\lambda}),h_2(\boldsymbol{\lambda}))$, satisfying Equation~(\ref{eq:criteria_full}) is a {\it sufficient} condition for the standard indistinguishability criterion of Equation~(\ref{eq:FF}) to hold in realistic, $D$-dimensional data analysis. Equation~(\ref{eq:criteria_full}) therefore imposes a more stringent (conservative) constraint on waveform validity.

We use the x-model to generate PN waveforms in this work \citep[see, e.g.,][]{Hinder+10}. The x-model is a time-domain, PN-based waveform family that incorporates features introduced by eccentricity in non-spinning binaries \citep{Huerta+14}. In this approach, the binary dynamics are described through a Keplerian parameterization up to 3PN order, with the conservative evolution also given to 3PN. The gravitational wave energy and angular momentum losses are mapped to the evolution of the orbital eccentricity $e$ and the PN expansion parameter $x \equiv (\omega M)^{2/3}$, where $\omega$ denotes the mean Keplerian orbital frequency. These quantities are evolved using equations up to 2PN. This model has been verified against numerical relativity simulations for equal-mass BBHs with moderate eccentricity ($e=0.1$), showing good agreement over the final $\sim$21 inspiral cycles before merger. In the circular limit, it is also consistent with standard waveform templates commonly used in gravitational wave data analysis \citep[e.g.,][]{Brown+10}.

To match the realistic observation, we set $T_{\rm obs}=1$\,yr and compute the overlaps between time-domain PN signals generated with the x-model. Specifically, for a given set of source parameters, we generate two waveforms: waveform 1, denoted as $h_{\rm 2PN}$, constructed using 2PN conservative orbital dynamics and 1PN radiation reaction; and waveform 2, denoted as $h_{\rm 3PN}$, constructed using 3PN conservative dynamics and 2PN radiation reaction (see Appendix of \citet{Hinder+10} for details). We then compute the overlap $\mathcal{O}(h_{\rm 2PN}, h_{\rm 3PN})$ in frequency domain (Equation~\ref{eq:overlap}), and estimate the critical signal-to-noise ratio, $\rm SNR_c$, for $h_{\rm 2PN}, h_{\rm 3PN}$ to be indistinguishable in data analysis (see Equation~(\ref{eq:criteria_full}):
\begin{equation}
{\rm SNR_c}=
\sqrt{\frac{D-1}{2[1-\mathcal{O}(h_{\rm 2PN}, h_{\rm 3PN})]}}\ ,
\label{eq:snr_c}
\end{equation}
where $D$ equals 10 for eccentric, non-spinning BBHs \citep{Xuan24parameter}.

In general, $\rm SNR_c$ represents the SNR threshold below which PN expansion converges well (thus may be safe to use). For example, $\rm SNR_c = 100$ implies that the difference between 2PN and 3PN waveforms is indistinguishable for the GW sources' parameter inference, provided that the detected signals have SNR below 100.

\begin{figure*}
    \centering
    \includegraphics[width=7in]{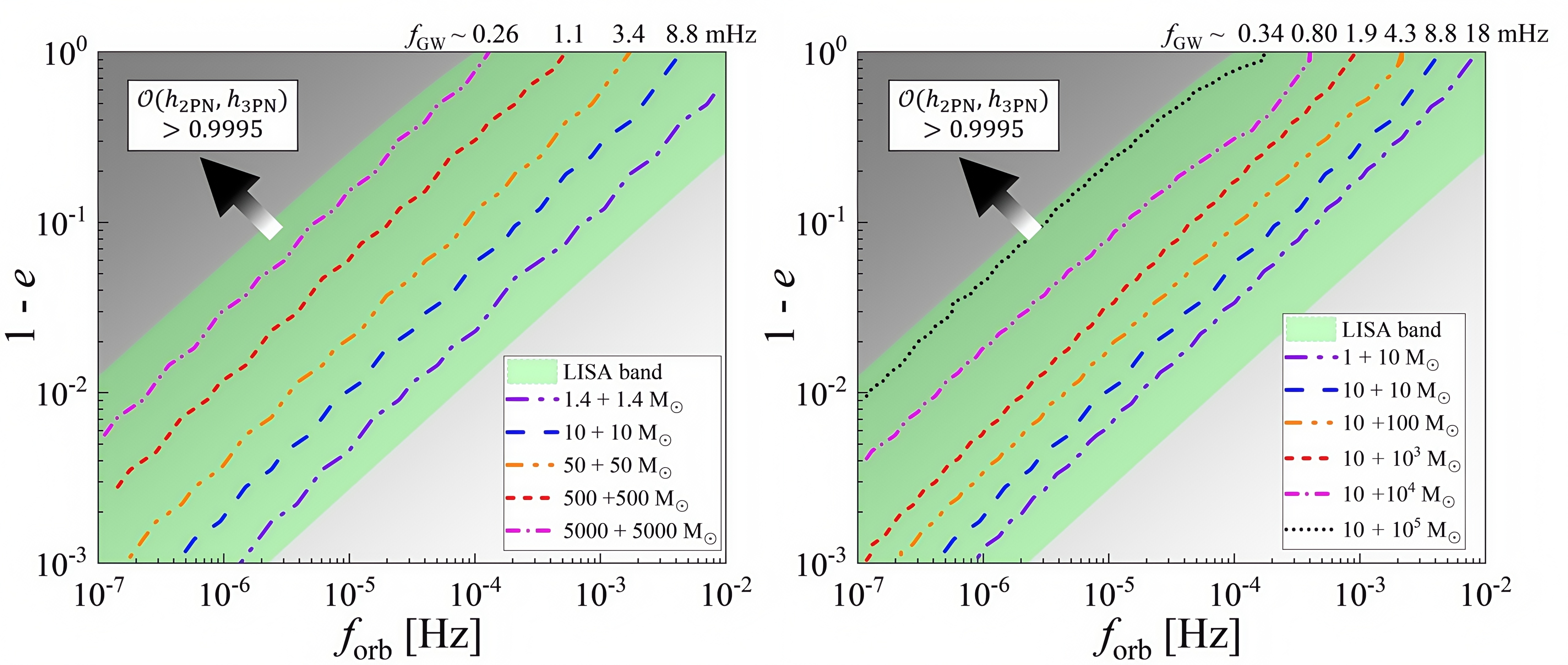}
    \caption{\textbf{Regime of validity for PN waveforms with different  binary orbital frequencies, eccentricities, and masses.} We plot the boundary of parameter space where 2PN and 3PN waveforms converge well in data analysis ($\mathcal{O}(h_{\rm 2PN}, h_{\rm 3PN}) > 0.9995$, corresponding to $\mathrm{SNR}_c > 100$) over a 1-year LISA observation. {\it Left Panel} shows the boundaries for equal-mass binaries ranging from $1.4+1.4\,M_\odot$ to $5000+5000\,M_\odot$. {\it Right Panel} shows unequal-mass systems ranging from $1+10\,M_\odot$ to $10+10^5\,M_\odot$. For a given mass, the region to the upper-left of the dashed contour, as indicated by the dark grey arrow, represents the high-accuracy regime where the PN waveforms are safe to use. The green shaded area represents the LISA sensitive band ($0.1$~mHz -- $0.1$~Hz). We note that the minor ``wiggles'' along the contours are computational artifacts caused by numerical fluctuations and the discrete parameter grid sampling in the overlap integrals (see Equation~\ref{eq:overlap}).}
    \label{fig:pn_accuracy2}
\end{figure*}

Figure~\ref{fig:pn_accuracy} shows the resulting $\mathrm{SNR}_c$, as estimated via numerical waveform generation and overlap computation. In particular, the {\it Left Panel} displays the color map of $\mathrm{SNR}_c$ as a function of the periastron distance $r_{\rm p}$ and eccentricity $e$, assuming a fixed BBH mass of $20+25\,M_\odot$. For reference, we plot the contours of constant peak GW frequencies $f_{\rm GW} = 1\,\mathrm{mHz}$ and $10\,\mathrm{mHz}$ (dashed lines), and use red hatches to shade the region where $\mathrm{SNR}_c \lesssim 100$, indicating that the 2PN and 3PN waveforms fail to converge. The primary takeaway from the {\it Left Panel} is that for stellar-mass binaries, the validity of the PN approximation is mostly determined by the peak GW frequency at pericenter, with little dependency on the orbital eccentricity. As long as the GW signal remains below $f_{\rm GW} \sim 10\,\mathrm{mHz}$, $\mathrm{SNR}_c$ can safely exceed $100$. This indicates that PN waveforms can be reliable for analyzing eccentric, including highly eccentric, stellar-mass BBH signals in the mHz band.

The {\it Right Panel} shows $\mathrm{SNR}_c$ as a function of the total binary mass $M $ and peak frequency $f_{\rm GW}$, assuming an equal-mass binary with a fixed eccentricity of $e=0.9$. The solid gray region in the bottom-right corner is excluded from the analysis because the periastron distance in this regime is smaller than the radius of the innermost stable circular orbit ($r_{\rm p} < r_{\rm ISCO}$). As demonstrated, the convergence of the PN expansion exhibits a strong dependence on the total mass. While stellar-mass binaries converge well in the typical mHz band, the $\mathrm{SNR}_c \gtrsim 100$ threshold shifts to lower frequencies as the mass grows. Therefore, for massive and supermassive black hole binaries, PN waveforms are only reliable for low-frequency GW data analysis (e.g., $f_{\rm GW}\sim 0.1-1$~mHz). Additionally, we test the convergence of PN expansion for different mass ratios by fixing the secondary mass at $m_2 = 10\,M_\odot$ and increasing the primary mass $m_1$. It turns out that the $f_{\rm GW}$ threshold for PN convergence behaves similarly to the equal mass binary cases (Figure~\ref{fig:pn_accuracy}, {\it Right Panel}), generally agreeing within a factor of two.

Figure~\ref{fig:pn_accuracy2} shows a generalized view of the PN waveform convergence test, across the parameter space of orbital frequency, eccentricity, and mass. In particular, we select a set of representative compact object masses (see the legend in Figure~\ref{fig:pn_accuracy2}). For each mass, the $\rm SNR_c$ distribution is computed and mapped following a procedure similar to Figure~\ref{fig:pn_accuracy}. Here, however, the distribution is mapped as a function of $f_{\rm orb}$ and $1 - e$, instead of $e$ and $r_p$. We then extract the equal-$\rm SNR_c$ contour corresponding to $\rm SNR_c = 100$ in the map, which defines the boundary of the region where PN waveforms converge sufficiently well (i.e., $\mathcal{O}(h_{\rm 2PN}, h_{\rm 3PN})>0.9995$, see Equation~\ref{eq:snr_c}). By varying the total mass of the binary, we track how this high-accuracy boundary shifts across the $f_{\rm orb}$–$(1 - e)$ parameter space, as shown by the dashed lines in Figure~\ref{fig:pn_accuracy2}.

{\it Left Panel} of Figure~\ref{fig:pn_accuracy2} represents the cases of equal-mass binaries, while {\it Right Panel} shows unequal-mass systems. For comparison, we highlight the region corresponding to the LISA band ($0.1\,\mathrm{mHz} < f_{\rm GW} < 0.1\,\mathrm{Hz}$, green shaded band), and mark on the top axis the GW frequency corresponding to each dashed line in the circular limit. As shown, the high accuracy region of PN waveforms spans a broad range of parameter space (toward the upper-left side of each contour). As the mass of binaries increases, this highly PN accuracy region gradually shrinks and shifts toward lower frequencies, eventually moving below the LISA band. 

This behavior effectively sets an upper limit of GW frequency for each binary mass, above which PN waveforms are no longer reliable in data analysis. However, as suggested by Figure~\ref{fig:pn_accuracy2}, PN methods can safely cover most of mHz sources with stellar mass. Specifically, the high accuracy region extends up to $f_{\rm GW} \sim 9\,\mathrm{mHz}$ for a $10+10\,M_\odot$ BBH, even at extreme eccentricity values (provided that $f_{\rm GW} \sim 2f_{\rm orb}(1-e)^{-3/2}\lesssim 9\,\mathrm{mHz}$). Additionally, PN methods may still be useful for analyzing massive systems at lower GW frequencies, such as $\sim 1\,\mathrm{mHz}$ signals from intermediate-mass black hole binaries ($M\sim 1000\,M_\odot$), and $\sim 0.3\,\mathrm{mHz}$ signals from extreme mass ratio inspirals ($M\sim 10^5\,M_\odot$).

\section{Discussion}
\label{sec:discussion}

Eccentric stellar-mass BBHs arise naturally in dynamical formation environments and can contribute significantly to mHz gravitational wave observations \citep[see, e.g.,][]{Fang19, Xuan+23b, Xuan_2025gc}. Their signals exhibit distinctive waveform features, characterized by ``repeated bursts" near periapsis \citep{Kocsis_2012, Loutrel+20, Xuan+23b,Romero23,knee2024detectinggravitationalwaveburstsblack}. In this work, we present a simulated catalog of BBHs formed via dynamical channels, characterizing their population properties in both the Milky Way and at cosmological distances. We further evaluate the impact of these eccentric sources on the LISA global fit, assess the validity of corresponding PN waveform templates, and provide a ready-to-use software package for their population and waveform analysis.

Specifically, we introduce the GW radiation and evolutionary properties of eccentric binaries in Section~\ref{subsec:basicproperty} (see Figure~\ref{fig:burst_waveform} and Table~\ref{tab:typical_params}), then discuss different conventions in the literature to plot their characteristic strain spectrum (Section~\ref{subsec:hcreps}, Figure~\ref{fig:hc_representations}). Based on recent dynamical simulations, we constructed a mock catalog in Section~\ref{subsec:catalog}, which incorporates contributions from the Galactic field \citep[fly-by induced mergers,][]{Michaely+19, Michaely+22}, Galactic nuclei \citep[EKL induced mergers,][]{Hoang+18, Xuan+23b}, and globular clusters \citep[Monte-Carlo N-body results,][]{Kremer_2020, Xuan_2025gc}. Our study shows that a significant number of eccentric BBHs will be detectable in the Milky Way. \xzy{For a 4-year (10-year) LISA observation, approximately $25, 8.5, 3.0, 1.5, 0.4$ ($36, 13, 4.7, 2.3, 1.0$) dynamically-formed MW BBHs can exceed SNR thresholds of $1, 3, 8, 20, 50$, respectively (see Table~\ref{tab:snr_stats}).} 

Because of their dynamical origins, these BBHs exhibit distinct yet universally scaled parameter and SNR distributions. In particular, their evolutionary tracks trace a characteristic pathway in the $a$–$(1-e)$ plane: systems begin on wide orbits and undergo eccentricity excitation to very high $e$ (with negligible change in $a$), until gravitational wave emission decouples them from their dynamical environments. This transition appears as a sharp feature in the $a$–$(1-e)$ plane, followed by GW-dominated orbital shrinkage and circularization (see Figure~\ref{fig:mock_mw}). 

Meanwhile, different formation channels can exhibit distinct distributions in semi-major axis and eccentricity. For example, the Galactic field channel typically produces wider BBHs ($a \sim 1$–$10$~au) with more extreme eccentricities ($e \sim 0.999$) upon entering the LISA band, whereas globular cluster and Galactic nucleus channels yield more compact systems ($a \sim 0.1$~au) with moderately high eccentricities ($e \sim 0.7$–$0.9$). Interestingly, our simulations suggest a similar number of detectable sources for different channels (Field, GC, and GN) in high SNR thresholds (up to a factor of a few). However, these channels do exhibit a different power-law SNR distribution (Figure~\ref{fig:snrdistribution}), which results in different low-SNR BBH populations. Furthermore, despite their specific population distribution, detectable mHz BBHs in the MW all reside along a narrow, elongated region in the $a-(1-e)$ parameter space (see, e.g., Figures~\ref{fig:mock_mw}, \ref{fig:mw_bbh_distribution}, and \ref{fig:orb-dl}), which roughly corresponds to a GW merger timescale of $t_{\rm merger}\sim 10^7$~yrs and marks the region where GW radiation decouples the system from the dynamical environment.

We further expanded this analysis to extragalactic sources in Section~\ref{subsec:bbhcosmos}, evaluating both their merger eccentricities in the LVK band (Figure~\ref{fig:ecc_cdf}) and their orbital parameters in the LISA band (Figure~\ref{fig:orb-dl}). Our results show that the observable BBHs are dominated by two major populations: wide, highly eccentric BBHs in the local universe ($D_l \sim 10\,\mathrm{kpc}$, median $e \sim 0.9$) that produce mHz repeated GW bursts; and quasi-circular, LIGO-type inspirals at cosmological distances ($D_l \sim 100 - 1000\,\mathrm{Mpc}$; median $e \sim 0.1 - 0.2$ in mHz band; with varying LVK merger eccentricities, see Figure~\ref{fig:ecc_cdf}). While the extragalactic population contributes significantly at low-SNR regime ($\sim 150$ sources with $\mathrm{SNR} > 1$), this number drops drastically at higher thresholds in mHz GW detection ($\sim 6$ for $\mathrm{SNR} > 3$, and $\lesssim 1$ for $\mathrm{SNR} > 8$), making them a largely faint background compared to the loud local GW bursts. 

Notably, eccentric BBHs will significantly affect the LISA global fit. In Section~\ref{sec:globalfit}, we demonstrated that individual harmonics from the local eccentric BBH population can be independently detected. Specifically, we expect an average of \xzy{$\sim 216,\, 60,\, 10,\, \text{and } 3$} individual harmonics with $\mathrm{SNR} > 1, 3, 8, 20$, respectively (Figure~\ref{fig:SNR_HARM}). These harmonics can be misidentified as circular signals from binaries with biased, systematically smaller chirp masses (Figures~\ref{fig:hcscatter} and \ref{fig:fakemc}). Furthermore, our estimation of the stochastic background indicates that the collective background from Milky Way highly eccentric BBHs sits roughly one order of magnitude below the nominal LISA noise curve (Figure~\ref{fig:hc}). However, combining it with the extragalactic cosmological background could produce a confusion noise that may interfere with the detection of other mHz sources \citep[see][for details]{Xuan24bkg}.

We further test the accuracy of PN waveforms for the data analysis of eccentric stellar-mass BBHs in LISA. Specifically, we used numerical waveform generation and inner product overlaps to map the convergence regions of PN-based templates (Figures~\ref{fig:pn_accuracy} and \ref{fig:pn_accuracy2}). We found that the PN expansion generally converges well for eccentric stellar-mass BBH detection, provided that the system has a peak GW frequency $f_{\rm GW}\lesssim 10\,\mathrm{mHz}$. Furthermore, the PN method could be leveraged for lower-frequency searches of heavier sources, and remains accurate for binary masses up to $\sim 10^3\,M_\odot$ (with $f_{\rm GW}\lesssim 1\,\mathrm{mHz}$). However, in both low and high eccentricity cases, the PN method has strict limitations; the upper boundary of its valid frequency range is typically insufficient to support signal processing near the LISA high-frequency limit ($\sim 0.1\,\mathrm{Hz}$), or for sources with SMBH components (e.g., eccentric extreme mass ratio inspirals). This highlights the need for waveform modeling techniques, such as post-Minkowskian expansion \citep[e.g.,][]{Bern_2019, Khalil_2022}, self-force approach \citep[e.g.,][]{Barack_2018}, and numerical relativity \citep[e.g.,][]{Pretorius_2005, Baker_2006}.

\xzy{The mock catalog presented here should be viewed as a conservative baseline rather than a complete representation of the Galactic BBH population. In particular, it does not include binaries formed through isolated binary evolution, nor other potentially important dynamical channels such as mergers in AGN accretion disks or isolated hierarchical triples in the field.} Nevertheless, our results, based on recent simulations and observational evidence \citep{Hoang+19,Kremer_2020,Michaely+20,Xuan+23b,Xuan_2025gc}, provide a ready-to-use population model for LISA data challenges and global fit studies. Moreover, since eccentricity is a generic outcome of dynamical formation, there exist other potential eccentric GW populations, such as dynamically-formed BH-WD or BH-NS binaries \citep[e.g.,][]{xuan2025UCXB}, as well as eccentric extreme mass ratio inspirals \citep[e.g.,][]{Amaro_Seoane_2019}. While these sources may exhibit different orbital features, they are expected to share the characteristic burst-like GW signatures and highly eccentric evolutionary behavior highlighted in this work.

The catalog generation and waveform modeling developed in this work are implemented in an open-source Python package, the \textit{LISA Eccentricity Astrophysics Package} (\href{https://github.com/zeyuanxuan/lisa-leap/}{\texttt{LEAP}}). This package provides mock catalogs, along with PN-based time-domain waveform generation and data analysis tools tailored for eccentric systems, enabling these bursting sources to be accurately modeled and fully leveraged in the upcoming era of space-based GW astronomy.

\acknowledgments
\xzy{The authors thank the anonymous referee for their useful comments, and thank Pau Amaro Seoane, Yonadav Barry Ginat, and Mor Rozner for their valuable discussions.} ZX acknowledges partial support from the Bhaumik Institute for Theoretical Physics summer fellowship.
ZX and SN acknowledge partial support from NASA ATP Grant No.~80NSSC24K0773, 
and thank Howard and Astrid Preston for their generous support. MLK's research was carried out at the Jet Propulsion Laboratory, California Institute of Technology, under a contract with the National Aeronautics and Space Administration (80NM0018D0004).

\bibliographystyle{apsrev4-2.bst}
\bibliography{bibbase}

\appendix
\section{Eccentric GW signal analysis}
\label{sec:appendixa}

\subsection{Analytical SNR Calculation}
For completeness, here we summarize and refine the derivations from previous works \citep[see, e.g., Appendix B of][]{Xuan+23b} regarding the signal-to-noise ratio calculation of eccentric GW sources.

We first present the analytical calculation of the sky-averaged SNR. The time-domain waveform of an eccentric binary, $h(a, e, t)$, can be decomposed into different harmonics, with the frequency of the $n$-th harmonic given by $f_n = n f_{\rm orb}$ \citep{peters63,Kocsis_2012_hcn}:
\begin{equation}
h(a, e, t) = \sum_{n=1}^{\infty} h_n(a, e, f_n) \exp (2 \pi i f_n t),
\end{equation}
where $h(a,e,t)$ represents the complex waveform combining both polarizations. The corresponding effective real waveform can be written as $h(t)_{\rm real}=\sum_{n=1}^{\infty} \sqrt{2}\,h_n(a, e, f_n) \cos(2\pi f_n t + \phi_n)$, where $\phi_n$ is the initial phase of the $n$-th harmonic. Furthermore, the harmonic amplitude is given by:
\begin{equation}
h_n(a, e, f_n) = \frac{2}{n} \sqrt{g(n, e)} h_0(a)
\label{eq:hntime}
\end{equation}
which represents the inclination-averaged, time-domain root-mean-square (rms) strain amplitude of the $n$-th harmonic. This quantity incorporates the combined contributions from both GW polarizations. In addition, the overall amplitude $h_0(a)$ is defined as:
\begin{equation}\label{eq:hn}
h_0(a) = \sqrt{\frac{32}{5}} \frac{m_1 m_2}{D_l a}\,,
\end{equation}
and $g(n, e)$ is given by:
\begin{equation}\label{eq:gne}
\begin{split}
g(n, e) &= \frac{n^4}{32} \Bigg[ \left(J_{n-2} - 2e J_{n-1} + \frac{2}{n} J_n + 2e J_{n+1} - J_{n+2}\right)^2 \\
&\quad + (1-e^2)\left(J_{n-2} - 2 J_n + J_{n+2}\right)^2 + \frac{4}{3 n^2} J_n^2 \Bigg],
\end{split}
\end{equation}
where $J_i$ is the $i$-th Bessel function of the first kind evaluated at $ne$. 

The SNR of a given GW waveform can be computed via an integral in the frequency domain \citep[see, e.g.,][]{Robson+19,chen19}:
\begin{equation}\label{eq:snrintegral}
{\rm SNR}^2(a, e) = \int \frac{h_c^2(a, e, f)}{f^2 S_n(f)} d f,
\end{equation}
where $h_c$ is the characteristic strain and $S_n(f)$ is the effective noise power spectral density
of the detector, weighted by the sky and polarization-averaged signal response function of the instrument. We note that the subscript ``n'' in $S_n(f)$ stands for ``noise'', which is different from the harmonic index $n$ used throughout this section.

For an eccentric GW source, its characteristic strain $h_c(f)$ peaks at each harmonic's frequency, and its amplitude at the $n$-th harmonic, $h_{c,n}$, can be estimated as \citep{Finn_2000}:
\begin{equation}\label{eq:hcn}
h^2_{c,n}  = h_n^2 \cdot \frac{2f_n^2}{\dot{f}_n},
\end{equation}
where $\dot{f}_n = n \cdot \dot{f}_{\rm orb}$ is the time derivative of the $n$-th harmonic frequency, driven by the intrinsic orbital decay of the binary. A more detailed discussion of characteristic strain will be presented in Appendix~\ref{app:hc}. Also, note that here we take the pre-integral (fundamental) definition of $h_{c,n}$, see Equation~(\ref{eq:hcmixed}) for details.

As Equation~(\ref{eq:hcn}) shows, the value of $h_c$ is inversely proportional to the binary's orbital frequency shift. In the limit of a non-evolving source (i.e., $\dot{f}_n \rightarrow 0$), the value of $h^2_{c,n}$ in Equation~(\ref{eq:hcn}) diverges mathematically. This divergence reflects the fact that a perfectly monochromatic signal would deposit its energy into an infinitesimally narrow frequency bin $\Delta f_n$ over an infinite observation time; however, it is an artifact of the stationary phase approximation (SPA) underlying Equation~(\ref{eq:hcn}), and does {\it not} imply a divergent SNR. As we show next, evaluating the frequency integral in Equation~(\ref{eq:snrintegral}) yields a per-harmonic contribution that is independent of $\dot{f}_n$ (Equation~\ref{eq:snrsum}). Moreover, for genuinely non-evolving sources ($\dot{f}_n = 0$), the finiteness of the SNR is guaranteed independently by the post-integral characteristic strain (Equation~\ref{eq:hcnonevolve}) together with Parseval's theorem, rather than by the formal cancellation shown below; the two routes nonetheless yield consistent final results.

Specifically, assuming the binary has an orbital chirping rate $\dot{f}_{\rm orb}$ over an observation time $\Delta T_{\rm obs}$, the frequency shift of its $n$-th GW harmonic can be estimated as:
\begin{equation}
\Delta f_n = \dot{f}_{n} \Delta T_{\rm obs} = n\dot{f}_{\rm orb} \Delta T_{\rm obs}\,,
\label{eq:fndot}
\end{equation}
which effectively represents the frequency width of the $n$-th harmonic peak. 
Assuming $S_n(f)$ is roughly constant across this narrow bin, we can sum the contributions of each harmonic peak to the integral in Equation~(\ref{eq:snrintegral}):
\begin{equation}\label{eq:snrsum0}
\begin{aligned}
{\rm SNR}^2(a, e) &\approx \sum_n \int_{f_n}^{f_n+\Delta f_n} \frac{h_c^2(a, e, f)}{f^2 S_n(f)} d f \\
&\approx \sum_n \frac{h_{c,n}^2}{f_n^2 S_n(f_n)} \Delta f_n .
\end{aligned}
\end{equation}
Plugging in Equations~(\ref{eq:hcn}) and (\ref{eq:fndot}), we get:
\begin{equation}\label{eq:snrsum}
\begin{aligned}
{\rm SNR}^2(a, e) &\approx  \sum_n \frac{h_{c,n}^2}{f_n^2 S_n(f_n)} \Delta f_n \\
&= \sum_n \frac{h_n^2 \cdot \left(2f_n^2/\dot{f}_n\right)}{f_n^2 S_n(f_n)} \cdot \dot{f}_n \Delta T_{\rm obs} \\
&= \sum_n \frac{2h_n^2}{S_n(f_n)} \Delta T_{\rm obs}\, .
\end{aligned}
\end{equation}
Equation~(\ref{eq:snrsum}) essentially demonstrates that the total squared SNR (power) of an eccentric GW signal is the sum of the squared SNRs of its individual harmonics, where each harmonic is mathematically treated as an independent circular GW source \citep[see, e.g., eq. 35 in][]{Moore_2014}. A visual representation of this discrete harmonic structure is shown in Figure~\ref{fig:burst_waveform}.

We note that the cancellation of $\dot{f}_n$ between Equations~(\ref{eq:hcn}) and (\ref{eq:fndot}) is carried out in the evolving regime, where $\Delta f_n \gg 1/T_{\rm obs}$ and the SPA underlying Equation~(\ref{eq:hcn}) holds. The resulting per-harmonic contribution, $2h_n^2 \Delta T_{\rm obs}/S_n$, is itself independent of $\dot{f}_n$, and therefore extends smoothly to the non-evolving (long-lived) sources discussed below. In that limit, the same result follows directly from the post-integral characteristic strain (Equations~\ref{eq:hcnonevolve} and \ref{eq:snrnonevolve}) and is independently guaranteed by Parseval's theorem (Appendix~\ref{app:snrnumerical}), without invoking the SPA. This agreement between the two routes is what justifies applying Equation~(\ref{eq:snrsum2}) to long-lived sources, as we do next.

Specifically, substitute $h_n$ from Equation~(\ref{eq:hn}) into Equation~(\ref{eq:snrsum}), and get:
\begin{equation}
    {\rm SNR}^2 = 8h^2_0(a) \sum_n \frac{g(n,e)}{S_n(n f_{\rm orb})n^2} \Delta T_{\rm obs} \ .
    \label{eq:snrsum2}
\end{equation}

Equation~(\ref{eq:snrsum2}) can be directly applied to the SNR estimation of long-lived GW sources, assuming that the evolution of the orbital parameters is negligible during $\Delta T_{\rm obs}$. In a more general case, however, the system's parameters ($a, e, f_{\rm orb}$) evolve as functions of time during the observation window. Therefore, the result in Equation~(\ref{eq:snrsum2}) should be treated as the differential SNR contribution (${\rm SNR}^2 \rightarrow d{\rm SNR}^2$) over an infinitesimal observation window ($\Delta T_{\rm obs} \rightarrow dt$). The total SNR over a finite observation time, ${\rm SNR}_{\rm tot}$, is then obtained by integration:
\begin{align}
    {\rm SNR}_{\rm tot}^2 &= \int_0^{T_{\rm obs}} d{\rm SNR}^2(t) \nonumber \\
    &= \int_0^{T_{\rm obs}} 8h^2_0(a(t)) \sum_n \frac{g(n,e(t))}{S_n(n f_{\rm orb}(t))\,n^2} dt \,.
    \label{eq:snrsum3}
\end{align}
Equation~(\ref{eq:snrsum3}) provides the most general SNR formulation for evolving eccentric GW sources.

Additionally, when the binary's orbit is highly eccentric, calculating the sum over millions of harmonics in Equations~(\ref{eq:snrsum2}) and (\ref{eq:snrsum3}) becomes computationally expensive. However, we can accelerate this mode summation by utilizing the continuous envelope of the frequency spectrum. In other words, instead of summing over millions of discrete harmonics, we can estimate their average amplitude within a wider frequency bin and multiply by the density of peaks below that envelope, thereby approximating the SNR via a smoothed frequency integration.

Consider a fixed frequency $f$ and a narrow integration interval $df$ around it. The number of harmonic peaks contained within this interval is proportional to the bin width, $dn = df/f_{\rm orb}$ (note that the separation between harmonics is $f_{\rm orb}$), and the average amplitude of these peaks can be estimated using a local average of $g(n_{\rm avg}, e)$ in Equation~(\ref{eq:snrsum3}), where $n_{\rm avg} = f/f_{\rm orb}$. 

Thus, the discrete mode summation in Equation~(\ref{eq:snrsum2}) transforms into a continuous frequency integral:
\begin{equation}\label{eq:snrsum4}
\begin{aligned}
    {\rm SNR}^2 &\approx 8h^2_0(a)\Delta T_{\rm obs} \int \frac{g(n_{\rm avg},e)}{S_n(f)n_{\rm avg}^2} dn \\
    &= 8h^2_0(a)\Delta T_{\rm obs} \int \frac{g(n_{\rm avg},e)}{S_n(f)\left(\frac{f}{f_{\rm orb}}\right)^2} \frac{df}{f_{\rm orb}} \\
    &= 8h^2_0(a) f_{\rm orb} \Delta T_{\rm obs} \int \frac{g(n_{\rm avg},e)}{S_n(f) f^2} df \ ,
\end{aligned}
\end{equation}
and similarly for Equation~(\ref{eq:snrsum3}):
\begin{equation}
    {\rm SNR}_{\rm tot}^2 \approx \int_0^{T_{\rm obs}} 8h^2_0(a(t))f_{\rm orb}(t) \left[ \int \frac{g(n_{\rm avg},e(t))}{S_n(f)f^2} df \right] dt \ .
    \label{eq:snrsum5}
\end{equation}
Equations~(\ref{eq:snrsum4}) and (\ref{eq:snrsum5}) serve as fast numerical estimators for the SNR of highly eccentric binaries. The accuracy of the integral can be controlled by adjusting the step size $df$. In the limit of $df = f_{\rm orb}$, the strict discrete summation of Equations~(\ref{eq:snrsum2}) and (\ref{eq:snrsum3}) is recovered.

\begin{figure*}
    \centering
    \includegraphics[width=0.47\textwidth]{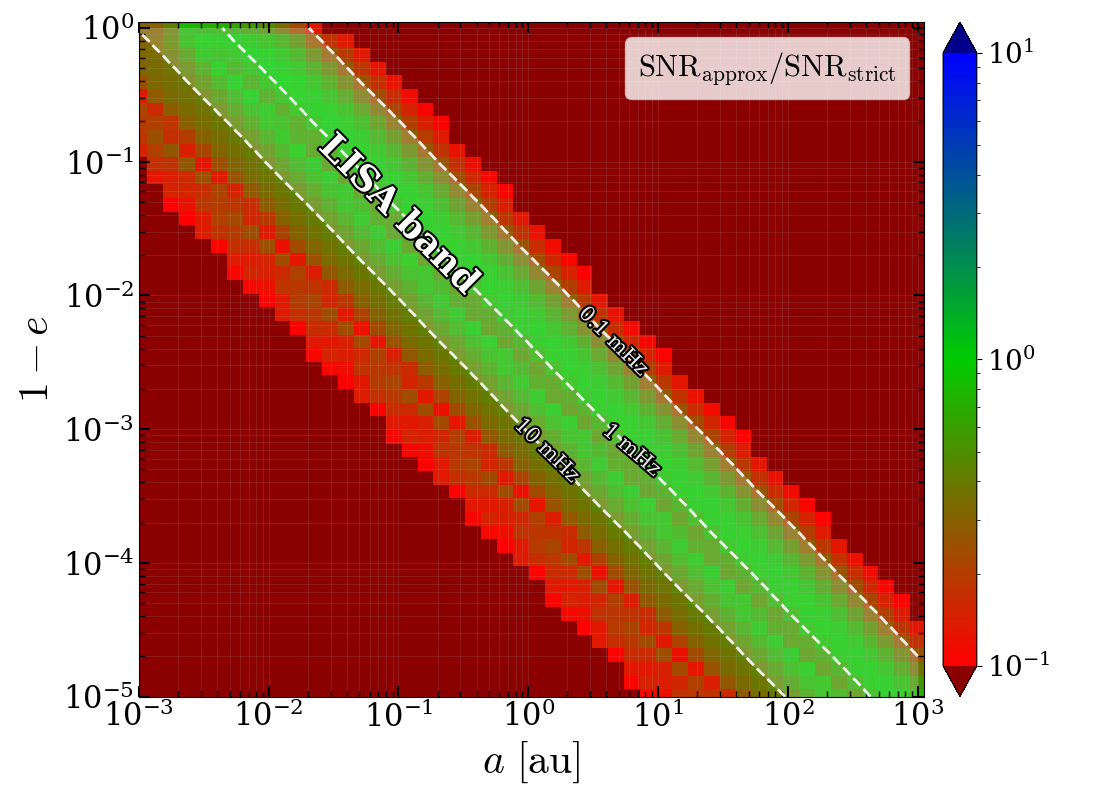}
    \includegraphics[width=0.47\textwidth]{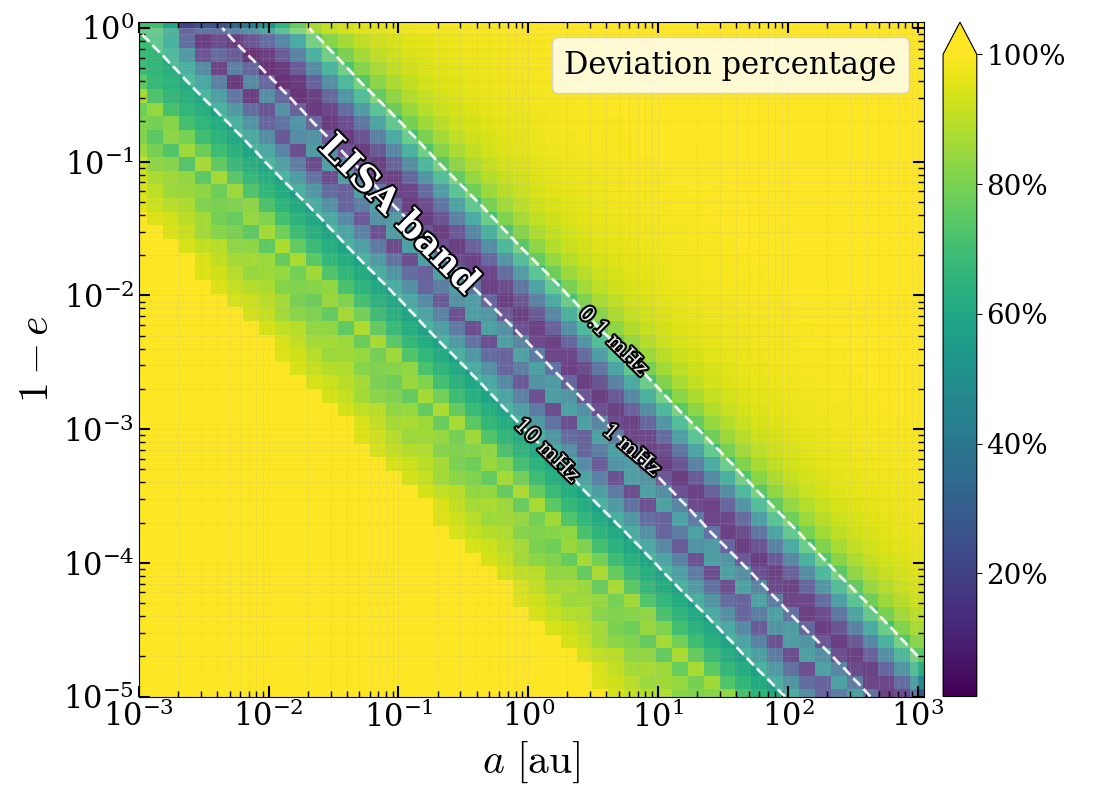}
\caption{\textbf{Comparison between different SNR calculation methods.} Here we compute the analytical SNR approximation, $\rm SNR_{approx}$, Equation~(\ref{eq:snrnew}), and the strictly computed SNR, $\rm SNR_{strict}$, Equation~(\ref{eq:snrsum3}) for a $10+10$~$M_{\odot}$ binary with different orbital parameters. {\it Left panel:} The numerical ratio of $\rm SNR_{approx}$ to $\rm SNR_{strict}$. The green region highlights where the ratio is close to unity. {\it Right panel:} The relative deviation percentage between the two calculations. The region bounded by the white dashed lines represents the LISA sensitivity band ($0.1 \rm{\, mHz}$ to $10 \rm{\, mHz}$). Within this band, the analytical approximation is consistent with the strict calculation, with the relative deviation generally below $60\%$.}
    \label{fig:pn_accuracy3}
\end{figure*}

In addition, to validate the analytical approximation discussed in Section~\ref{sec:burst_properties}, we compare the strictly calculated SNR from Equation~(\ref{eq:snrsum3}) with the approximate analytical estimation given by Equation~(\ref{eq:snrnew}). The comparison across the parameter space is illustrated in Figure~\ref{fig:pn_accuracy3}. As shown in the figure, both methods yield generally consistent results within the LISA band. The ratio of the approximated SNR to the strict SNR ($\rm SNR_{approx}/SNR_{strict}$) is distributed closely around unity, and the relative deviation percentage is below $60\%$ for the majority of the parameter space. This consistency confirms that Equation~(\ref{eq:snrnew}) is a reliable SNR estimation for general BBH population studies within the LISA band.

\subsection{Analytical Characteristic Strain Calculation}
\label{app:hc}

In the gravitational wave literature, the definition and computation of the characteristic strain, $h_c$, often depend on the observational context and the evolutionary stage of the source. These varying conventions can introduce ambiguities when comparing sensitivity curves with source amplitudes. In this section, we clarify the conventions adopted in this work and relate them to the literature.

For an inspiraling source with GW frequency derivative $\dot{f}$ and time-domain rms strain amplitude $h_0$, the characteristic strain is defined as \citep{Finn_2000,Moore_2014}
\begin{equation}\label{eq:hcinsp}
h_{c, \rm insp} = h_0 \sqrt{\frac{2f^2}{\dot{f}}} \ .
\end{equation}
This definition has a clear physical interpretation: as the signal evolves, the area enclosed by the $h_c$ evolution track and the noise curve in a log-log plot is directly proportional to the square of the signal-to-noise ratio \citep[see eq.~19 in][]{Moore_2014}.

However, for non-evolving or slowly evolving sources such as Galactic binaries, their GW frequency shift is negligible during observation ($\dot{f} \to 0$). Applying the previous definition would cause $h_c$ to diverge, resulting in an infinitely high and narrow track in the characteristic strain plot (since an infinite time to evolve across a frequency bin implies an infinite SNR accumulation). In reality, the observation time $T_{\rm obs}$ is finite, which always yields a finite SNR. Therefore, it is conventional to plot the $h_c$ of slowly evolving systems using an alternative definition \citep[see, e.g., ][]{Kupfer_2018,Tauris_2018,Robson+19}:
\begin{equation}\label{eq:hcnonevolve}
    h_{c, \rm non-evolve} = \sqrt{2} h_0 \sqrt{N} = \sqrt{2} h_0 \sqrt{f T_{\rm obs}} \ ,
\end{equation}
meaning the characteristic strain of a monochromatic signal is proportional to the square root of the number of cycles, $N = f T_{\rm obs}$, that the signal undergoes during the observation. The additional factor of $\sqrt{2}$ accounts for the fact that our $h_0$ is defined as the time-averaged rms amplitude rather than the peak amplitude. This alternative definition plots the signals as discrete points in the characteristic strain plot, with the height of each point representing their post-integration SNR.

In the literature, the definitions in Equations~(\ref{eq:hcinsp}) and (\ref{eq:hcnonevolve}) are frequently mixed or combined into a unified formula. Adapted for our rms amplitude convention, this becomes:
\begin{equation}\label{eq:hcmixed}
    h_c = \sqrt{2} h_0 \sqrt{ \min\left\{ \frac{f^2}{\dot{f}}, \, f T_{\rm obs} \right\} } \ .
\end{equation}

However, the physical meanings of the two regimes in Equation~(\ref{eq:hcmixed}) are fundamentally different: The $\dot{f}$-dependent characteristic strain (Equation~\ref{eq:hcinsp}) is a \textit{pre-integral} quantity, which is related to the SNR via the standard frequency integral (Equation~\ref{eq:snrintegral}); In contrast, the $T_{\rm obs}$-dependent expression (Equation~\ref{eq:hcnonevolve}) is a \textit{post-integral} quantity, for which the SNR is evaluated directly without performing the frequency integral:
\begin{equation}
    {\rm SNR}^2 = \frac{h_{c, \rm non-evolve}^2}{f S_n(f)} \ . \label{eq:snrnonevolve}
\end{equation}

This distinction is reflected in the characteristic strain plot: for evolving sources with $f^2/\dot{f}<fT_{\rm obs}$, the height of their characteristic strain curve is independent of $T_{\rm obs}$, while $T_{\rm obs}$ determines the evolution time for $h_{c, \rm insp}$, thus affecting the enclosed area and accumulated SNR. Conversely, for non-evolving sources, the height of the point depends on $T_{\rm obs}$, and its ratio to the detector noise directly represents the accumulated SNR over $T_{\rm obs}$.

\subsection{Plotting Characteristic Strain of (Highly) Eccentric Systems}\label{app:plothighe}

There are different approaches to visualizing the characteristic strain of (highly) eccentric sources, each serving a different analytical purpose. 

The first approach, which we refer to as the {\it individual harmonic representation}, plots each harmonic independently using the $h_c$ definitions in Equation~(\ref{eq:hcn,0}). However, for highly eccentric sources, the number of harmonics is large, and their frequency evolution is often slow in the mHz band ($\dot{f} \to 0$). As a result, plotting individual harmonics can become overwhelmingly cluttered, and hard to reflect the overall detectability of an eccentric source (see, e.g., the scatter points in Figure~\ref{fig:hcscatter}).

The second approach, which we refer to as the {\it smoothed spectrum representation}, introduces a smoothed characteristic strain envelope, $h_{c,\rm env}(f)$, for a highly eccentric GW source. This approach is designed to capture the cumulative spectral contribution of a single eccentric binary across different GW frequencies (see Equation~\ref{eq:hc_highe0}). 

Specifically, consider a fixed frequency $f$ and a narrow interval $df$ around it. The number of harmonics contained within this interval is $dn = df / f_{\rm orb}$. Treating each harmonic as a monochromatic, slowly evolving source, its contribution to the SNR can be estimated using Equations~(\ref{eq:hcnonevolve}) and (\ref{eq:snrnonevolve}):
\begin{equation}
    \Delta {\rm SNR}_{\rm single}^2 \sim \frac{h_{c,n}^2(f)}{f S_n(f)} = \frac{(\sqrt{2}h_n(f) \sqrt{f T_{\rm obs}})^2}{f S_n(f)} = \frac{2h_n^2(f) T_{\rm obs}}{S_n(f)} \ ,
\end{equation}
where $h_{c,n}(f)=\sqrt{2} h_n(f) \sqrt{f T_{\rm obs}}$ is the averaged characteristic strain for harmonics near frequency $f$, $h_n(f)=2/n_{\rm avg}\cdot \sqrt{g(n_{\rm avg}, e)} h_0(a)$ and $n_{\rm avg}=f/f_{\rm orb}$, see Equations~(\ref{eq:hntime}) - (\ref{eq:gne}). See also Equations~(\ref{eq:snrsum4}) - (\ref{eq:snrsum5}) for a similar derivation.

Therefore, the total contribution to the signal-to-noise ratio from all harmonics within a frequency interval $df$ around $f$ is
\begin{equation}\label{eq:snrsumhighe}
d{\rm SNR}^2 = dn \times \Delta {\rm SNR}_{\rm single}^2 = \frac{df}{f_{\rm orb}} \frac{2 h_n^2(f) T_{\rm obs}}{S_n(f)} \ .
\end{equation}
where $dn$ denotes the number of harmonics within $df$.

We then define the continuous characteristic strain envelope as:
\begin{equation}\label{eq:hc_highe}
h_{c,\rm env}(f) = \sqrt{2}\, h_n(f)\,  \sqrt{\frac{f^2T_{\rm obs}}{f_{\rm orb}}} \ ,
\end{equation}


This definition ensures that the standard characteristic strain integral of $h_{c, \rm env}$ (Equation~\ref{eq:snrintegral}) over a frequency bin $df$ reproduces the correct accumulated SNR from all harmonic peaks within the same bin, as computed previously in Equation~(\ref{eq:snrsumhighe}): \begin{equation}\label{eq:highehc_snr} \int d{\rm SNR}^2  = \int \frac{df}{f_{\rm orb}} \frac{2 h_n^2(f) T_{\rm obs}}{S_n(f)} = \int \frac{h_{c, \rm env}^2(f)}{f^2 S_n(f)}\, df \ . \end{equation}

We note that $h_{c,\rm env}$ does not correspond to the geometric upper envelope of individual harmonic tracks in Equation~(\ref{eq:hcmixed}). Instead, its physical interpretation is that, on a log--log characteristic strain plot, the area between $h_{c,\rm env}(f)$ and $\sqrt{fS_n(f)}$ is directly related to the SNR. This allows one to visually estimate the detectability of the source \citep{Moore_2014}, as well as the GW energy contribution from different frequency ranges (see, e.g., the blue curves in Figure~\ref{fig:hc}).

Finally, the third approach, which we refer to as the {\it numerical spectrum representation}, is used in numerical waveform analysis, where the time-domain waveform with finite length is transformed to a spectrum via a Fast Fourier Transform (FFT). This yields $h_{c,\text{num}}[k]$, as described below in Appendix~\ref{app:hcnumerical}.

While the individual harmonic ($h_{c,n}$), smoothed spectrum ($h_{c,\rm env}(f)$), and numerical spectrum ($h_{c,\text{num}}[k]$) representations describe the same underlying physical signal, they differ fundamentally in their construction and in how the total SNR is computed. In the {\it individual harmonic representation}, each harmonic in the eccentric signal is treated as a quasi-circular source. Their $\rm SNR^2$ is computed using Equation~(\ref{eq:snrintegral}) when $f^2/\dot{f} < f T_{\rm obs}$, and Equation~(\ref{eq:snrnonevolve}) when $f^2/\dot{f} > f T_{\rm obs}$, then summed together to yield the total $\rm SNR^2$ of the eccentric source; In the {\it smoothed spectrum representation}, $h_{c,\rm env}$ renormalize discrete harmonic power over frequency, and the total SNR is obtained via a continuous integral (Equation~\ref{eq:highehc_snr}); In contrast, {\it the numerical spectrum} $h_{c,\text{num}}$ can apply to arbitrary time-domain waveforms, and its SNR is computed via a direct, discrete summation (Equation~\ref{eq:snrnum-hc}). Although all three approaches converge to the same total SNR for highly eccentric systems, their visual representations on the $h_c$ plot can differ significantly. See Figure~\ref{fig:hc_representations} for a comparison.

\subsection{Numerical SNR Calculation}
\label{app:snrnumerical}
Consider a discrete time-domain signal $h_t[i] = h(i \cdot t_s)$, where $i = 0, 1, \dots, N-1$, $t_s$ is the sample interval, and the total observation time is $T_{\text{obs}} = N t_s$. We assume the total number of samples $N$ is an even integer for simplicity.

Applying a one-sided Fast Fourier Transform to $h_t[i]$ yields the complex frequency-domain array $H[k]$ (for the positive frequency components). To extract the physical time-domain amplitude, we apply the normalization:
\begin{equation}
\sqrt{2}h_{\rm FFT}[k] = |H[k]| \frac{2}{N}\, ,
\end{equation}
where $h_{\rm FFT}[k]$ is the rms time-domain amplitude of the $k$-th discrete Fourier component at frequency $f[k]$ (which equals $1/\sqrt{2}$ times of the maximum time-domain amplitude); $|H[k]|$ is the norm of the complex array $H[k]$; the factor of $2/N$ arises because of the numerical property of FFT and the fact that we only take the positive frequency components. While the full one-sided spectrum spans $k=0, 1, \dots, N/2$ (yielding $N/2+1$ discrete bins with $\Delta f=1/T_{\rm obs}$), in practical numerical implementations, both the DC component ($k=0$) and the Nyquist component ($k=N/2$) are dropped. This prevents singular noise power at zero frequency and potential high-frequency artifacts.

With the corresponding phase $\phi[k]$ of $H[k]$, the time-domain signal can be reconstructed as a sum of monochromatic components at frequencies $f[k]$:
\begin{equation}
h_t[i] \sim \sum_{k} \sqrt{2}h_{\rm FFT}[k] \cos(2\pi f[k] i t_s + \phi[k])\, .
\end{equation}
This decomposition holds for an arbitrary signal: $h_{\rm FFT}[k]$ is simply the amplitude of the $k$-th component on the FFT frequency grid required to reconstruct $h_t[i]$, without any assumption about the underlying signal form.

Note that any realistic signal has a finite duration, which introduces non-trivial effects in the numerical results of $h_{\rm FFT}[k]$. In particular, a finite observation window of duration $T_{\rm obs}$ is equivalent to multiplying the underlying signal by a rectangular window in the time domain, which corresponds to convolving the true spectrum with a sinc-shaped kernel of width $\sim 1/T_{\rm obs}$ in the frequency domain. For the multi-harmonic signals considered in this work, $h_{\rm FFT}[k]$ recovers the true harmonic amplitude $h_n$ at $f[k]=nf_{\rm orb}$ only when every harmonic frequency $nf_{\rm orb}$ coincides exactly with an FFT grid point (i.e., when $nf_{\rm orb}T_{\rm obs}$ is an integer for all relevant $n$); in that case, the sinc kernel's zeros happen to fall on all neighboring grid points, leaving $h_{\rm FFT}[k]$ non-zero only at the harmonic bins. In the more general case, where the harmonic frequencies do not align with the FFT grid, each harmonic's power can leak into neighboring bins, so that $h_{\rm FFT}[k]$ stays non-zero even when $f[k]$ does not equal the harmonic frequencies $nf_{\rm orb}$; and the sampled peak at $f[k]\approx nf_{\rm orb}$ falls slightly below $h_n$. Nevertheless, Parseval's theorem guarantees that the total signal power is preserved across all bins, so that summing $|h_{\rm FFT}[k]|^2$ over the bins into which a given harmonic leaks recovers the true power $|h_n|^2$ of that harmonic. Consequently, the bin-wise SNR sum constructed below correctly accounts for the leaked power and remains insensitive to the alignment between $nf_{\rm orb}$ and the FFT grid.

Specifically, treating $h_{\rm FFT}[k]$ at each frequency $f[k]$ as an independent, monochromatic signal within finite duration $T_{\text{obs}}$, its squared SNR contribution is (see Equation~\ref{eq:snrnonevolve}):
\begin{equation}
\text{snr}_k^2= \frac{(\sqrt{2}h_{\rm FFT}[k]\cdot \sqrt{ f[k]T_{\text{obs}}})^2}{f[k]\cdot S_n(f[k])} = \frac{2h_{\rm FFT}[k]^2 T_{\text{obs}}}{S_n(f[k])}\, ,
\end{equation}
where $S_n(f[k])$ is the one-sided noise Power Spectral Density, and the term $\sqrt{f[k]T_{\rm obs}}$ accounts for the number of cycles this frequency component contributes during $T_{\rm obs}$. The total $\text{SNR}^2$ for $h_t[i]$ waveform is simply the sum over the valid AC frequency bins: 
\begin{equation}
\text{SNR}^2 = \sum_{k=1}^{N/2-1} \text{snr}_k^2\, .
\end{equation}

\subsection{Numerical Characteristic Strain Calculation}
\label{app:hcnumerical}

For an arbitrary numerical waveform (such as those generated from highly eccentric binaries with complex harmonic structures), there is generally no \textit{a priori} analytical expression for the frequency derivative $\dot{f}$ at a given global frequency $f$. Consequently, the conventional analytical definition of the characteristic strain derived from the Stationary Phase Approximation, $h_c \propto \sqrt{f^2/\dot{f}}$, cannot be directly applied. 

Instead, we adopt the discrete time-domain representation $h_t[i]$. As established in the previous section, the FFT directly provides the discrete monochromatic physical amplitude $h_{\rm FFT}[k]$ at each frequency $f[k]$. For an individual frequency component observed over the total duration $T_{\text{obs}}$, the effective number of accumulated cycles is given by $N_{\text{eff}} = f[k] T_{\text{obs}}$. 

The numerical characteristic strain $h_{c, \text{num}}[k]$ at frequency $f[k]$ is then defined by scaling the discrete time-domain amplitude by the square root of the effective cycles:
\begin{equation}\label{eq:hnc_num}
h_{c, \text{num}}[k] = \sqrt{2}h_{\rm FFT}[k] \sqrt{N_{\text{eff}}} = \sqrt{2}h_{\rm FFT}[k] \sqrt{f[k] T_{\text{obs}}}\, .
\end{equation}

Note that this numerical formulation avoids the ambiguous $\dot{f}$ interpretation. The blue curves in Figure~\ref{fig:burst_waveform} are calculated following this method.

This definition guarantees mathematical consistency with the SNR calculation. By substituting $h_{c, \text{num}}[k]$ back into Equation~(\ref{eq:snrnonevolve}), we naturally recover the numerical SNR per bin:
\begin{align}\label{eq:snrnum-hc}
\text{snr}_k^2 &= \frac{h_{c, \text{num}}[k]^2}{f[k] S_n(f[k])} \nonumber \\
&= \frac{\left(\sqrt{2}h_{\rm FFT}[k] \sqrt{f[k] T_{\text{obs}}}\right)^2}{f[k] S_n(f[k])} = \frac{2h_{\rm FFT}[k]^2 T_{\text{obs}}}{S_n(f[k])}\, .
\end{align}
Equation~(\ref{eq:snrnum-hc}) demonstrates that defining $h_{c, \text{num}}[k] = \sqrt{2}h_{\rm FFT}[k] \sqrt{f[k] T_{\text{obs}}}$ correctly maps the raw FFT output to the physical sensitivity curve without requiring any assumptions about the orbital evolution rate.

\end{document}